\documentclass[fleqn,usenatbib]{mnras}

\usepackage[T1]{fontenc}

\usepackage{ae,aecompl}

\DeclareRobustCommand{\VAN}[3]{#2}
\let\VANthebibliography\thebibliography
\def\thebibliography{\DeclareRobustCommand{\VAN}[3]{##3}\VANthebibliography}

\usepackage{graphicx}	
\usepackage{amsmath}	
\usepackage{amssymb}	

\usepackage{ulem}
\usepackage{soul}
\usepackage{natbib}

\usepackage{xcolor}

\usepackage{lineno}

\usepackage{newtxtext,newtxmath}
\usepackage[T1]{fontenc}

\newcommand{\msun}{$\rm M_\odot$}

\definecolor{darkgreen}{rgb}{0.1, 0.6, 0.1}

\title[Dwarf galaxy morphology in FIREbox]{Disks no more: the morphology of low-mass simulated galaxies in FIREbox}

\author[J. A. Benavides et al.]{
Jos\'e A. Benavides$^{1}$\href{https://orcid.org/0000-0003-1896-0424}{\includegraphics[scale=0.8]{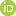}}\thanks{E-mail: jbenavid@ucr.edu},
Laura V. Sales$^{1}$\href{https://orcid.org/0000-0002-3790-720X}{\includegraphics[scale=0.8]{images/orcid.png}}, 
Andrew Wetzel$^{2}$\href{https://orcid.org/0000-0003-0603-8942}{\includegraphics[scale=0.8]{images/orcid.png}},
Jorge Moreno$^{3,4}$\href{https://orcid.org/0000-0002-3430-3232}{\includegraphics[scale=0.8]{images/orcid.png}},
Robert Feldmann$^{5}$\href{https://orcid.org/0000-0002-1109-1919}{\includegraphics[scale=0.8]{images/orcid.png}}, \newauthor
Francisco J. Mercado$^{6}$\href{https://orcid.org/0000-0002-5908-737X}{\includegraphics[scale=0.8]{images/orcid.png}},
James S. Bullock$^{7}$\href{https://orcid.org/0000-0003-4298-5082}{\includegraphics[scale=0.8]{images/orcid.png}},
Philip F. Hopkins$^{6}$\href{https://orcid.org/0000-0003-3729-1684}{\includegraphics[scale=0.8]{images/orcid.png}},
Claude-Andr\'e Faucher-Guig\`ere$^{8,9}$\href{https://orcid.org/0000-0002-4900-6628}{\includegraphics[scale=0.8]{images/orcid.png}}, \newauthor
Jonathan Stern$^{10}$\href{https://orcid.org/0000-0002-7541-9565}{\includegraphics[scale=0.8]{images/orcid.png}},
Coral Wheeler$^{11}$ and
Du\v{s}an Kere\v{s}$^{12,13}$\href{https://orcid.org/0000-0002-1666-7067}{\includegraphics[scale=0.8]{images/orcid.png}} \\
$^{1}$Department of Physics and Astronomy, University of California, Riverside, CA, 92507, USA\\ 
$^{2}$Department of Physics and Astronomy, University of California, Davis, CA 95616, USA\\ 
$^{3}$Department of Physics and Astronomy, Pomona College, Claremont, CA 91711, USA\\
$^{4}$Carnegie Observatories, Pasadena, CA 91101, USA\\
$^{5}$Department of Astrophysics, Universität Zürich, Zurich, CH-8057, Switzerland \\
$^{6}$TAPIR, California Institute of Technology, Pasadena, CA 91125, USA\\
$^{7}$Department of Physics and Astronomy, University of California, Irvine, CA 92697, USA\\
$^{8}$Center for Interdisciplinary Exploration and Research in Astrophysics (CIERA)\\
$^{9}$Department of Physics and Astronomy, Northwestern University, 1800 Sherman Avenue, Evanston, IL 60201, USA\\
$^{10}$School of Physics \& Astronomy, Tel Aviv University, Tel Aviv 69978, Israel\\
$^{11}$Department of Physics and Astronomy, California State Polytechnic University, Pomona, Pomona, CA 91768, USA \\ 
$^{12}$Department of Astronomy and Astrophysics, University of California, San Diego, La Jolla, CA 92093, USA \\ 
$^{13}$Department of Physics, University of California, San Diego, La Jolla, CA 92093, USA
} 

\date{Accepted XXX. Received YYY; in original form ZZZ}

\pubyear{2022}

\begin{document}
\label{firstpage}
\pagerange{\pageref{firstpage}--\pageref{lastpage}}
\maketitle

\begin{abstract}
	We study the morphology of hundreds of simulated central galaxies in the stellar mass range $M_\star=10^{7.5} \rm - 10^{11}~$\msun\, from the FIREbox cosmological volume. We demonstrate that FIREbox is able to predict a wide variety of morphologies, spanning from disk-dominated objects to spheroidal galaxies supported by stellar velocity dispersion. However, the simulations predict a strong relation between morphology (degree of rotational support) and stellar mass: galaxies comparable to the Milky Way are often disk-dominated while the presence of stellar disks mostly vanishes for dwarfs with $M_\star <10^9 ~$\msun. This defines a ``morphology transition'' regime for galaxies with $10^9 <M_\star/\rm{M_\odot}< 10^{10}$ in which disks become increasingly common, but below which disks are rare. We show that burstiness in the star formation history and the deepening of the gravitational potential strongly correlate in our simulations with this transition regime, with disks forming in objects with lower levels of burstiness in the last $\sim 6$ Gyr and halos with mass $\sim 10^{11} ~ \rm{M_{\odot}}$ and above. While observations support a transition towards thicker disks in the regime of dwarfs, our results are in partial disagreement with observations of at least some largely rotationally supported gas disks in dwarfs with $M_\star < 10^9$\msun. This study highlights dwarf morphology as a fundamental benchmark for testing future galaxy formation models.
\end{abstract}




\section{Introduction}

Successful formation of disks in cosmological numerical simulations of galaxies required striking a balance between strong stellar feedback and suppressed star formation efficiencies, particularly to prevent overproduction of stars at high redshifts \citep{Agertz2011,Guedes2011,Marinacci2014}. Several studies have highlighted that besides a quiet accretion history, where most stars are born in-situ, additional fundamental factors to the development of disky structures include avoiding the formation of stars from low angular-momentum gas \citep{Brook2011,Ubler2014} and coherence in the alignment of the angular momentum of the  accreted gas \citep{Scannapieco2009,Sales2012,GarrisonKimmel2018, Hafen2022}.\\

Almost unanimously, numerical codes have achieved these attributes by generating (or implementing) winds that help remove some of the gas that would otherwise inevitably cool from the halos into the galaxies. As numerical treatments to model the physics of the interstellar medium grew more complex and better resolved, predictions suggest that, in low mass halos, star formation tends to be ``bursty'', showing relatively large variations in the amount of stars formed in short periods of time \citep{Hopkins2014,FaucherGiguere2018,Bose2019}. This feature, while present in most simulations in the dwarfs regime, seems particularly  prominent in those subgrid models that attempt to follow the multi-phase nature of the gas (e.g., cooling below a temperature $T \leq 10^4$ K), as opposed to adopting an equation-of-state model to describe a hot/cold phase into a single fluid \citep[e.g., ][]{Springel2003}.\\

Efficient coupling of the feedback energy resulting from such bursty star formation may trigger vigorous outflows and galactic fountains \citep{Suarez2016, Martizzi2020,Orr2022, Barbani2023}, the formation of bubbles \citep{DallaVecchia2008, Keller2014, Kim2017, Chengzhe2024} and outflow/inflow cycles --or breathing mode-- \citep{ElBadry2016} which overall are disruptive to the coherence of angular momentum in the gas. In the case of cosmological simulations of Milky Way-like (MW-like) galaxies, what ultimately enables the settling and formation of disks is {\it the halo mass}. As dark matter halos grow, star formation shifts to less episodic and more continuous along with a deepening of the gravitational potential that ultimately confines the gas from outflows and facilitates the accretion of gas with coherent angular momentum cooling from the hot circum-galactic medium \citep[CGM, e.g., ][]{Sales2012,Clauwens2018,Hopkins2023,Stern2024}.\\

In the regime of dwarf galaxies ($M_{\star} < 10^9 ~ \rm{M_{\odot}}$), levels of star-formation, temperature of the CGM gas and halo masses may substantially differ from that in MW-like galaxies. This poses an interesting question for modern baryonic treatments that successfully reproduce disk formation in MW-like galaxies: what is the predicted morphology for galaxies of lower mass? The FIRE project aims at capturing in detail the physics of galaxy formation in sub-kpc scales \citep{Hopkins2018} which has successfully formed disks in MW-like halos \citep{GarrisonKimmel2018} along with a realistic population of satellite galaxies \citep{Wetzel2016,GarrisonKimmel2019}. While the model was initially developed for zoom-in cosmological simulations, morphology as a function of mass is a question better suited for full-volume cosmological simulations, which allow for the necessary sampling of environmental effects, halo mass function and diversity of assembly histories. FIREbox, which samples a $\sim$ 22 Mpc on-a-side cosmological volume using the FIRE baryonic treatment \citep{Feldmann2023} constitutes therefore a valuable tool to tackle galaxy structure in low mass galaxies in simulations that resolve the multi-phase nature of the interstellar gas.\\

Using galaxies from the IllustrisTNG model, \citet{Tacchella2019} found that dwarfs formed with a numerical model that correctly predict disk fractions for MW-mass galaxies ends up with less disk-dominated structure than their MW-mass counterparts. Moreover, a comparison with data from the Galaxy And Mass Assembly (GAMA) survey \citep{Baldry2010,Kelvin2012,Lange2016} suggests that the trend closely follows the morphological distribution of real galaxies in the $M_{\star} \sim 10^{9.5} \rm - 10^{11.5} ~ M_{\odot}$ range if one assumes that dwarf irregulars in the GAMA survey are dispersion-dominated. Limitations in the resolution of the simulations prevented to further explore the regime of even less massive dwarfs. More recently, \citet{Celiz2025} identified a clear stellar mass - morphology relation where low mass dwarfs are increasingly more dispersion-dominated using the higher-resolution TNG50 run \citep[see also ][]{Zeng2024}.\\

In this paper, we use the FIREbox simulation to study the morphology of galaxies predicted as a function of mass for a volume-complete sample simulated with a detailed ISM model that follows the multi-phase nature of the gas. Previous studies have hinted at the possibility that the {FIRE-2 model results in dwarf galaxies that are kinematically more dispersion-dominated than observed, for example, regarding the kinematics of HI gas \citep{El-Badry2018b}. However, these conclusions have primarily been based on a limited set of zoom-in simulations, rather than on a large-volume cosmological run. More recently, \citet{Klein2025} suggests that the shapes of dwarf galaxies in FIREbox are rounder than in observations, highlighting a potential tension between the observed morphology in this regime and predictions from this simulation. In this paper, we investigate the role of angular momentum support and  explore the physical mechanisms that give raise to a morphological transition with stellar mass. In particular, we concentrate on the halo and stellar mass scale where rotationally-supported structures transition from rare to most common, which for the FIREbox simulation occur at $M_{\star} \sim 10^{9} \rm - 10^{10} ~ M_{\odot}$.  Our paper is organized as follows. In Sec.~\ref{sec:sims} we briefly describe the simulation and some details of our sample of simulated galaxies. In Sec.~\ref{sec:main_results} we present our main results in the morphological transition and formation of the disk structure in different stellar mass ranges. In Sec~\ref{sec:m11b} we present the scenario of disk formation through a wet co-planar merger, as a possible path of disks formation in low-mass galaxies. Finally, we present a general discussion and conclusions in Sec.~\ref{sec:concl}.

\begin{figure*}
	\centering
	\includegraphics[width=1.0\textwidth]{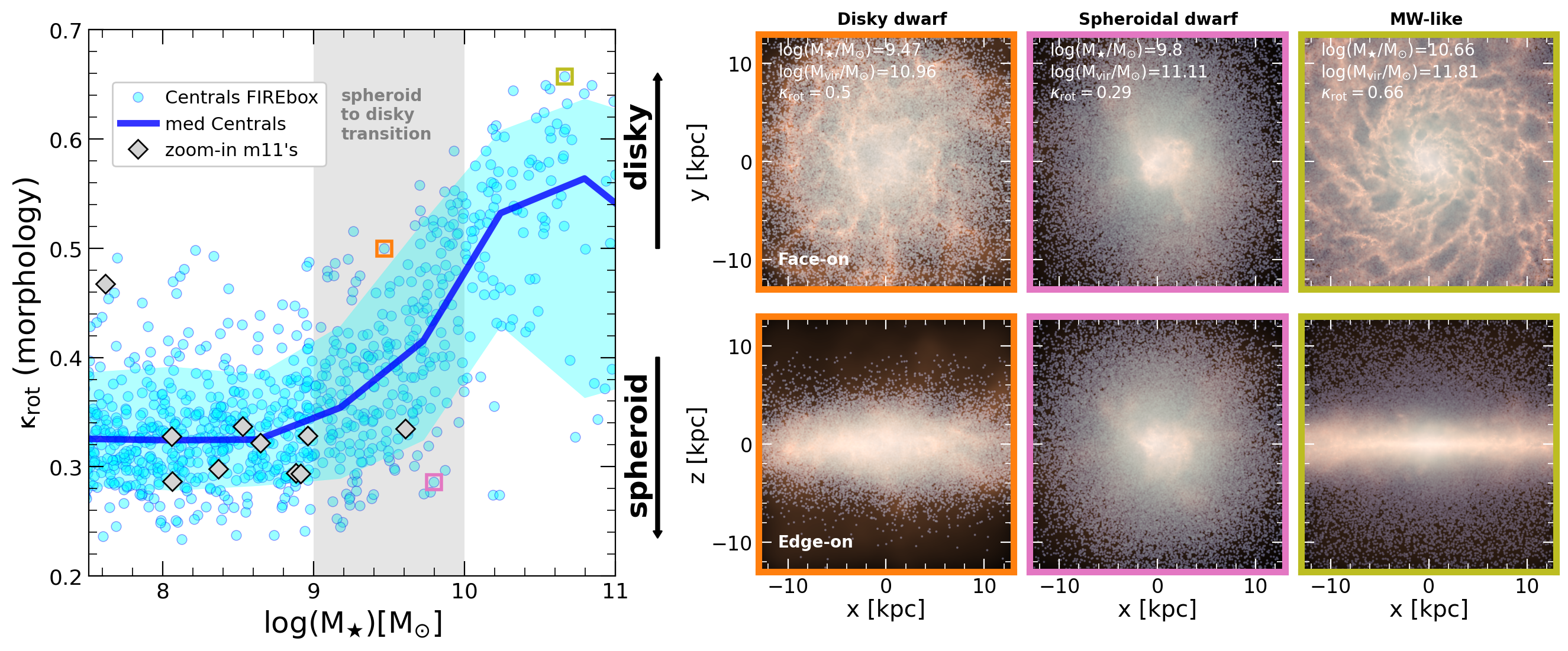}
	\caption{Left: Morphology ($\kappa_{\rm rot}$) as a function of stellar mass for all central galaxies in the FIREbox simulation (cyan circles). The median of the distribution is shown by the blue solid line, and the shaded cyan region encloses the 10th-90th percentiles. Morphology depends strongly on stellar mass: dwarfs are spheroidals while MW-like galaxies are mostly disk-dominated. We highlight in gray shading the ``transition region'' $M_{\star} = [10^9 - 10^{10}] ~ \rm{M_{\odot}}$ where disks begin to form to later dominate galaxy structure. Higher resolution \textit{m11} zoom-in runs are shown in gray diamonds, which agree well with the FIREbox morphologies. Right: examples of galaxies with varied morphology in FIREbox. The top and bottom rows show face-on and edge-on views of stars and gas for a disky dwarf (left), a spheroid-dominated dwarf (middle), and a disk-dominated MW-like galaxy (right). These objects are highlighted in color open squares on the left panel. The images were generated by using the Py-SPH Viewer library \citep{BenitezLlambay2017}}
	\label{fig:krot_Mstar}
\end{figure*}

\section{Simulations and method}
\label{sec:sims}

The simulations in this work were run using the FIRE-2 model \citep{Hopkins2018} coupled to the meshless ﬁnite-mass {\sc GIZMO} code \citep{Hopkins2015}. Gravity is calculated with a modiﬁed version of the
parallelization and tree gravity solver of GADGET-3 \citep{Springel2005}. The FIRE-2 model incorporates gas cooling down to $\rm{10 ~ K}$, which naturally leads to the formation of a multiphase interstellar medium (ISM). Once a gas cell becomes eligible for star formation, it converts to a star particle on a local free-fall time. Stellar feedback associated to radiation, winds, and supernova explosions is modelled by thermal or momentum input depending on whether the relevant scales are resolved \citep[see][for more details]{Hopkins2018b, Hopkins2020a}. The FIRE-2 model has been validated through numerous studies on galaxy formation and evolution, including satellite galaxy properties, MW-like morphologies, and numerical effect testing, all of particular relevance to this study \citep[see e.g.,][]{Wetzel2016, GarrisonKimmel2018, Hopkins2018}.\\

We use two different kinds of data from the Feedback In Realistic Environments (FIRE\footnote{\href{https://fire.northwestern.edu/}{https://fire.northwestern.edu/}}) project \citep{Hopkins2014, Hopkins2018}, i) FIREbox \citep{Feldmann2023} and ii) detailed individual zoom-in runs with halo masses $\sim 10^{11}$\msun\; (\textit{m11} runs) all reaching higher resolutions than FIREbox, as detailed in Table \ref{t:table1}. The cosmological parameters used in each simulation are slightly different and are described in their respective subsections. We briefly describe both data products below.

\subsection{FIREbox}
\label{ssec:firebox}
FIREbox follows the evolution of a 22.1 cMpc$^3$ volume set up initially with a total of $2 \times 1024^3$ of gas elements and dark matter particles. As all cosmological-volume simulations, it naturally resolves a myriad of environments and halo formation histories, offering a path to study galaxy properties and evolution with a statistical base.  The mass resolution of baryons is $m_{\rm bar} = 6.3 \times 10^4 ~ \rm{M_\odot}$ and $m_{\rm DM} = 3.3 \times 10^5 ~ \rm{M_\odot}$ for dark matter particles with a force resolution softening of 12 pc for stars and 80 pc for dark matter. The force softening of gas cells is adaptive, reaching a minimum of 1.5 pc in the dense ISM. FIREbox follows the evolution of multi-phase gas, modeled following the FIRE-2 framework. In this simulation, the initial conditions are generated at redshift $z = 120$ using the MUlti-Scale Initial Conditions \citep[MUSIC,][]{music2011} code. It assumes a set of cosmological parameters consistent with the \citet{Planck2016} measurements: $\Omega_{\rm m} = \Omega_{\rm dm} + \Omega_{\rm b} = 0.3089, \Omega_{\rm b} = 0.0486$, cosmological constant $ \Omega_{\rm \Lambda} = 0.6911$, Hubble constant $\rm{ H_0 = 100 \, h \, km \, s^{-1} \, Mpc^{-1} }$, $ h = 0.6774 $, $\sigma_8 = 0.8159 $ and spectral index $ n_s = 0.9667$. Previous studies of FIREbox focused on dwarf galaxies show the potential of FIRE-2 model this new large cosmological volume \citep{Moreno2022,Cenci2024,Klein2025,Mercado2025}.

\subsection{Individual low-mass galaxies zoom-in runs}
\label{ssec:m11_runs}
To explore the impact of numerical resolution, we compare FIREbox results with $10$ zoom-in runs from the FIRE project, with central host dark matter halos with masses of $M_{h} \sim 10^{11}~M_{\odot}$ (hereafter m11's, see Table~\ref{t:table1}). Note that the resolution per (baryonic) particle of all these zoom-in runs is $\sim 8$-$70$ $\times$ better than that achieved in FIREbox. In comparison with FIREbox, the zoom-in runs use a similar set of cosmological parameters, with small variations: $\Omega_{\rm m} = 0.266-0.31$, $\Omega_{\rm \Lambda} = 0.69-0.734$, $\Omega_{\rm b} = 0.044-0.048$, $\sigma_8 = 0.801-0.82$, and $n_s = 0.961-0.97$. However, at the scale of individual galaxies, variations due to differing cosmological parameters are small compared to run-to-run stochasticity. For details, please see the Public Data release \citep{Wetzel2023}. Also, small variations of the physics are included between runs, as follows: 

i) Metal diffusion (MD) and no-Metal diffusion (no MD), refer to an explicit treatment (or not) for the diffusion of metals \citep[e.g., ][]{Hopkins2017_MD,Hopkins2018}. ii) addition of cosmic rays (CR) feedback \citep{Chan2019_CR, Hopkins2020b}, which can drive multiphase winds, reduce star formation rates (mostly in low-mass galaxies) and modify the phase structure of the CGM, and iii) the higher resolution runs of \citet{El-Badry2018, Hopkins2018} but implementing the ultra-violet background (UVB) model of \citet{FaucherGiguere2020}, that permeates the intergalactic medium, keeping it ionized following the epoch of reionization. The other zoom-in used here implement the previous UVB model of \citet{FaucherGiguere2009}. These runs have all been introduced in several previous works \citep[see e.g.,][]{Chan2018, El-Badry2018, Hopkins2018, Hopkins2020b, Escala2018, Jahn2019, Hopkins2023}.

\begin{table}
	\centering
	\caption{Summary of simulations used in this work. First row correspond to our main sample from the FIREbox cosmological box. The subsequent lines describe the sample of high-resolution \textit{m11} zoom-in runs from the FIRE-2 suite. The first column: the simulation name, the second column: details of the physical model (no-MD: no metal diffusion, MD: with metal diffusion, MD+CR: metal diffusion and cosmic rays, uvb: Ultra-violet background), the third column: the initial resolution per gas cell in $\rm{M_\odot}$, the fourth and fifth column are the halo and stellar mass, respectively. The sixth column indicate the main reference of each simulation + their respective ultra-violet background model implemented: (1): \citet{Feldmann2023}, (2): \citet{Hopkins2018}, (3): \citet{El-Badry2018}, (4): \citet{Hopkins2020b}, $\dagger$: \citet{FaucherGiguere2009}, $\ddagger$: \citet{FaucherGiguere2020}}
	\setlength{\tabcolsep}{4pt}
	\begin{tabular}{cccccc}
		\hline\hline\noalign{\smallskip}
		\!\!\!Name & \!\!\!details &\!\!\!res. $m_b/\rm{M_{\odot}}$ & \!\!$M_{\rm vir}/\rm{M_{\odot}}$ & \!\! $M_\star/\rm{M_{\odot}}$ &  \! ref.\\
		\hline\noalign{\smallskip}
		FIREbox &  FIRE-2 &\!\!63000 &  $\sim 10^{9}-10^{12.4}$ & $10^{7.5}-10^{11}$ & (1)$\dagger$\\
		\hline\noalign{\smallskip}
		m11q &  MD &\!\!880 &  $1.3 \times 10^{11}$ & $3.4 \times 10^{8}$ & (2)$\dagger$ \\
		m11h &  UVB &\!\!880 &  $1.5 \times 10^{11}$ & $1.1 \times 10^{8}$ & (3)$\ddagger$\\
		m11q &  UVB &\!\!880 &  $1.2 \times 10^{11}$ & $2.3 \times 10^{8}$ & (2)$\ddagger$\\
		m11d &  UVB &\!\!880 &  $2.2 \times 10^{11}$ & $7.6 \times 10^{8}$ & (3)$\ddagger$\\
		m11e &  UVB &\!\!880 &  $1.3 \times 10^{11}$ & $4.4 \times 10^{8}$ & (3)$\ddagger$\\
		m11b &  MD+CR &\!\!2100 &  $3.9 \times 10^{10}$ & $4.2 \times 10^{7}$ & (4)$\dagger$\\
		m11a &  no-MD &\!\!2100 &  $4.1 \times 10^{10}$ & $1.1 \times 10^{8}$ & (2)$\dagger$\\
		m11c &  no-MD &\!\!2100 &  $1.4 \times 10^{11}$ & $8.2 \times 10^{8}$ & (2)$\dagger$\\
		m11d &  no-MD &\!\!7100 &  $3.2 \times 10^{11}$ & $4.0 \times 10^{9}$ & (4)$\dagger$\\
		m11i &  no-MD &\!\!7100 &  $7.8 \times 10^{10}$ & $9.1 \times 10^{8}$ & (4)$\dagger$\\
		\hline
	\end{tabular}
	\label{t:table1}
\end{table}

\subsection{Sample of simulated galaxies}
\label{ssec:sample}
We select all central galaxies from FIREbox (defined as the most massive centrally located galaxy in a dark matter halo) in the stellar mass range $M_{\star} = [10^{7.5} - 10^{11}] ~ \rm{M_{\odot}}$. The lower stellar mass bound represents $\sim 500$ stellar particles. From the \textit{m11} zoom-in runs, we select always the main (central) object only. No satellites are included in our samples. In FIREbox, halos and subhalos (galaxies) are identified using the AMIGA Halo Finder \citep[AHF,][]{AHF}. While, in the FIRE zoom-in sample, galaxies and their halos are identified using Rockstar \citep{rockstar}, with an extended algorithm to append the stellar and gas elements corresponding to each structure found in dark matter. Throughout this paper, virial quantities are measured using the radius within which the average density is 200 times the critical density of the Universe. Stellar masses are computed within the galaxy radius $r_{\rm gal}$, defined here as the radius that encloses 90\% of the stellar particles within the dark matter halo, associated to the central galaxy. The center of each object is defined as the most bound particle. In order to characterize the morphology of the simulated galaxies we use the $\kappa_{\rm rot}$ parameter \citep{Sales2012}, defined as a ratio that compares the energy in rotational support to the total kinetic energy of the stellar particles in a galaxy. More specifically,
\begin{equation}
\kappa_{\rm rot} = \frac{K_{\rm rot}}{K}=\frac{1}{K} \sum\;  \frac{1}{2} m \left( \frac{j_z}{R} \right)^2
\label{eq:krot}	\ \ ,
\end{equation}
where $K$ is the total kinetic energy of the galaxy calculated by using the stellar particles within the $r_{\rm gal}$, $j_z$ is the $z$-component of the angular momentum of each stellar particle (having rotated the galaxy, such that the total angular momentum points in the $z$ direction), $m$ denotes mass, $R$ is their cylindrical galactocentric distance and the sum is over stars within the galaxy radius. Typically, high values of $\kappa_{\rm rot} \geq 0.5$ are indicative of rotationally supported systems, such as disks, while lower values of $\kappa_{\rm rot} \leq 0.35$ are more characteristic of dispersion-dominated objects, such as traditional bulges. Intermediate values typically correspond to galaxies that contain both bulge and disk components or to dynamically hotter disks that are partially supported by dispersion.\\

The left panel of Fig.~\ref{fig:krot_Mstar} shows the $\kappa_{\rm rot} - M_\star$ relation of our sample of simulated galaxies. The cyan circles represent the sample of central galaxies from FIREbox, and the blue-thick solid line indicates the median $\kappa_{\rm rot}$ at a given stellar mass  (shadow region encloses the 10th-90th percentiles). Gray diamonds indicate the morphology of the zoom-in m11's runs. Encouragingly, FIREbox spans a much wider range of morphology than the selected zoom-in runs, highlighting the relevance of sampling different environments and assembly histories. We showcase three examples on the right panel of Fig.~\ref{fig:krot_Mstar}, selected to have different masses and morphologies (displayed from left to right): a disky and a spheroidal low-mass galaxies, and a MW-like disk-dominated galaxy. These FIREbox galaxies are highlighted on the left panel using colored open squares.\\

The left panel in Fig.~\ref{fig:krot_Mstar} highlights a clear trend predicted in the morphology of simulated galaxies: while disks are frequent and dominant for MW-like galaxies ($M_\star \geq 10^{10.5}~ \rm{M_{\odot}}$, large $\kappa_{\rm rot}$ values), they become progressively rarer for lower mass galaxies, to the point that no disks ($\kappa_{\rm rot} > 0.5$) are predicted in the simulation for dwarfs with $M_\star < 10^8~ \rm{M_{\odot}}$. This leads to a ``transition'' regime in the stellar mass range $M_\star \sim [10^9 - 10^{10}]~ \rm{M_{\odot}}$ (gray shaded region) where the frequency of rotationally-supported systems until they fully disappear below $M_\star \sim 10^9~ \rm{M_{\odot}}$. Note that the higher-resolution zoom-in runs agree with the expected median trend of FIREbox, suggesting that this trend is not driven by numerical resolution in the FIREbox simulation. We now investigate what factors play a role in this morphological transition in objects simulated with a fixed galaxy formation model that successfully reproduces healthy and observationally-compatible disks at scales of the MW.   

\begin{figure}
	\centering
	\includegraphics[width=\columnwidth] {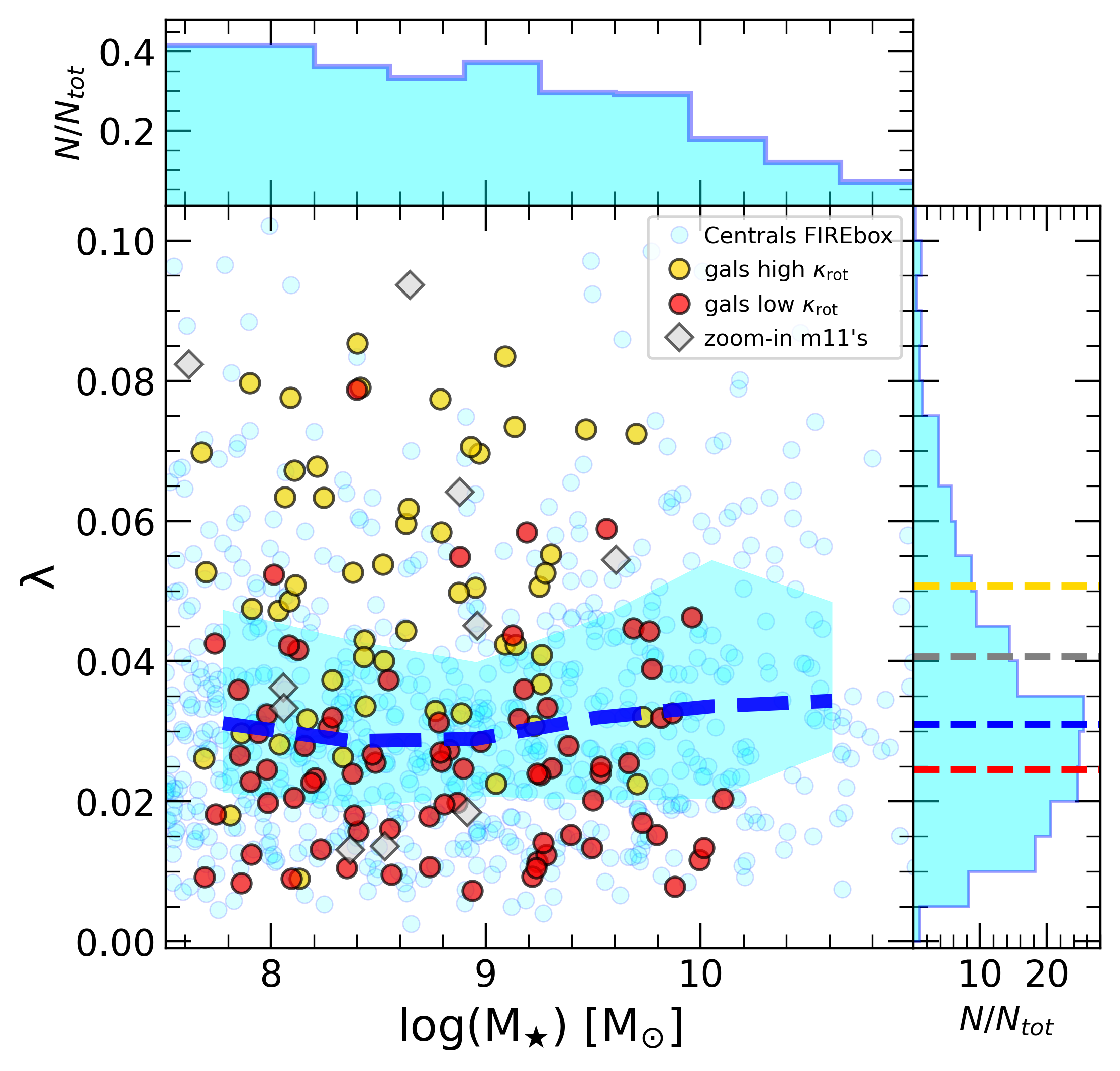}
	\caption{Halo dimensionless spin parameter ($\lambda$) as a function of stellar mass for all central galaxies in FIREbox simulation (cyan circles), along with the median in blue dashed and shaded regions enclose the 25th$-$75th percentiles. Yellow and red circles indicate the location of low-mass galaxies ($M_{\star} = [10^{7.5} - 10^{10}] ~ \rm{M_{\odot}}$) whose stars are most (high-$\kappa_{\rm rot}$) and least (low-$\kappa_{\rm rot}$) rotationally-supported, selected as the objects above and below the 90th and 10th percentile in the $\kappa_{\rm rot} - M_{\star}$ relation in Fig.~\ref{fig:examples}. The histograms in each axis represent the distribution of stellar mass and spin of the samples, along with their median $\lambda$ (colored dashed lines). The median (and the 25th$-$75th percentiles) of halo spin for all central galaxies is $\lambda_{\rm all}=0.031^{+0.014}_{-0.010}$, independent of $M_\star$ as expected. Morphology for low-mass dwarfs shows some dependence on the spin parameter, with high-$\kappa_{\rm rot}$ galaxies having $\lambda_{\rm{high}-\kappa_{\rm rot}}=0.051^{+0.019}_{-0.014}$ while the low-$\kappa_{\rm rot}$ subsample shows $\lambda_{\rm{low}-\kappa_{\rm rot}}=0.025^{+0.008}_{-0.009}$. We include the 10 zoom-in m11's FIRE runs with grey diamonds, with $\lambda_{\rm{m11's}}=0.041^{+0.021}_{-0.018}$, which agree with the overall FIREbox distribution.}
	\label{fig:lambda}
\end{figure}

\begin{figure*}
	\centering
	\includegraphics[width=0.325\textwidth]{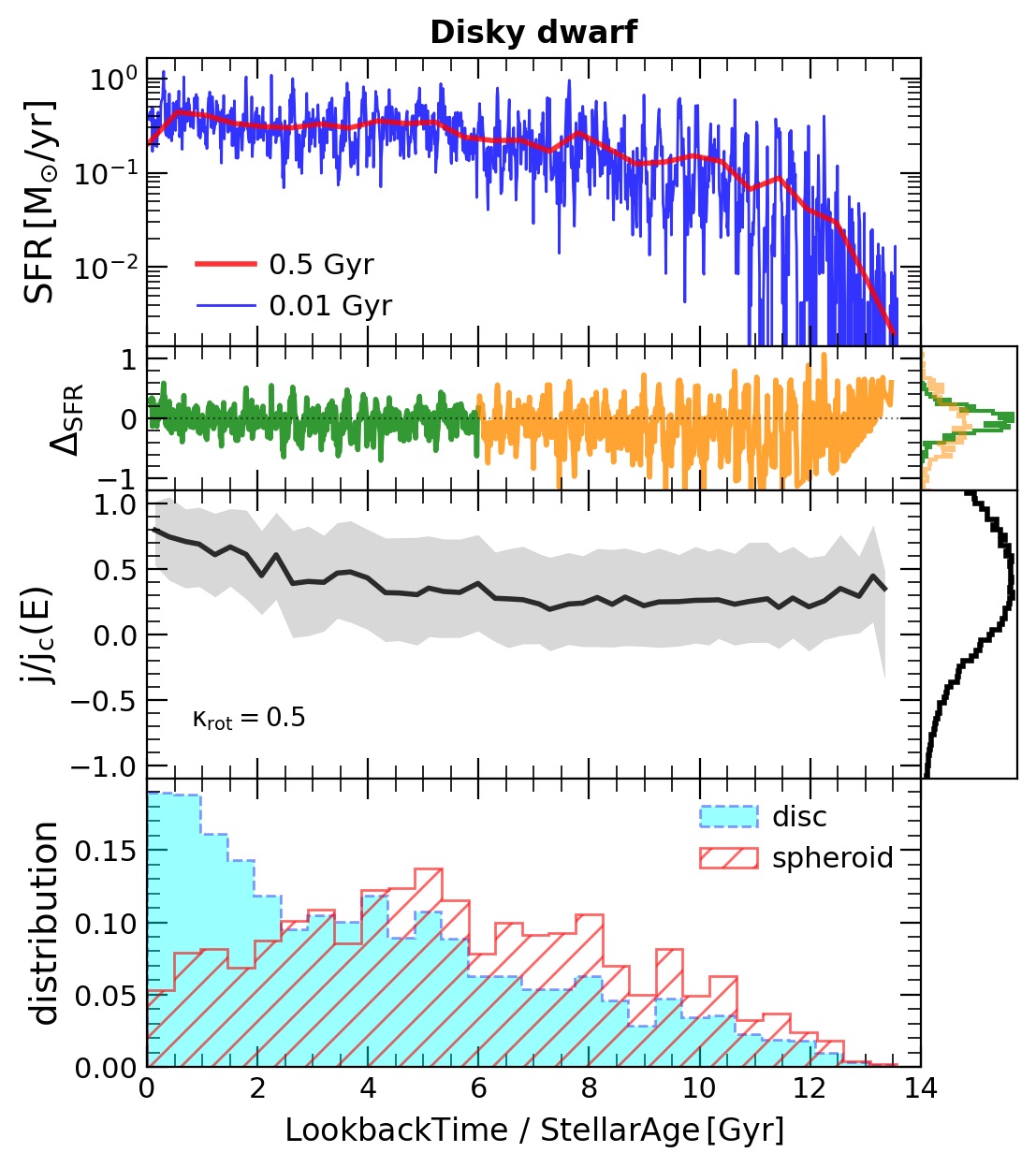}
	\includegraphics[width=0.325\textwidth]{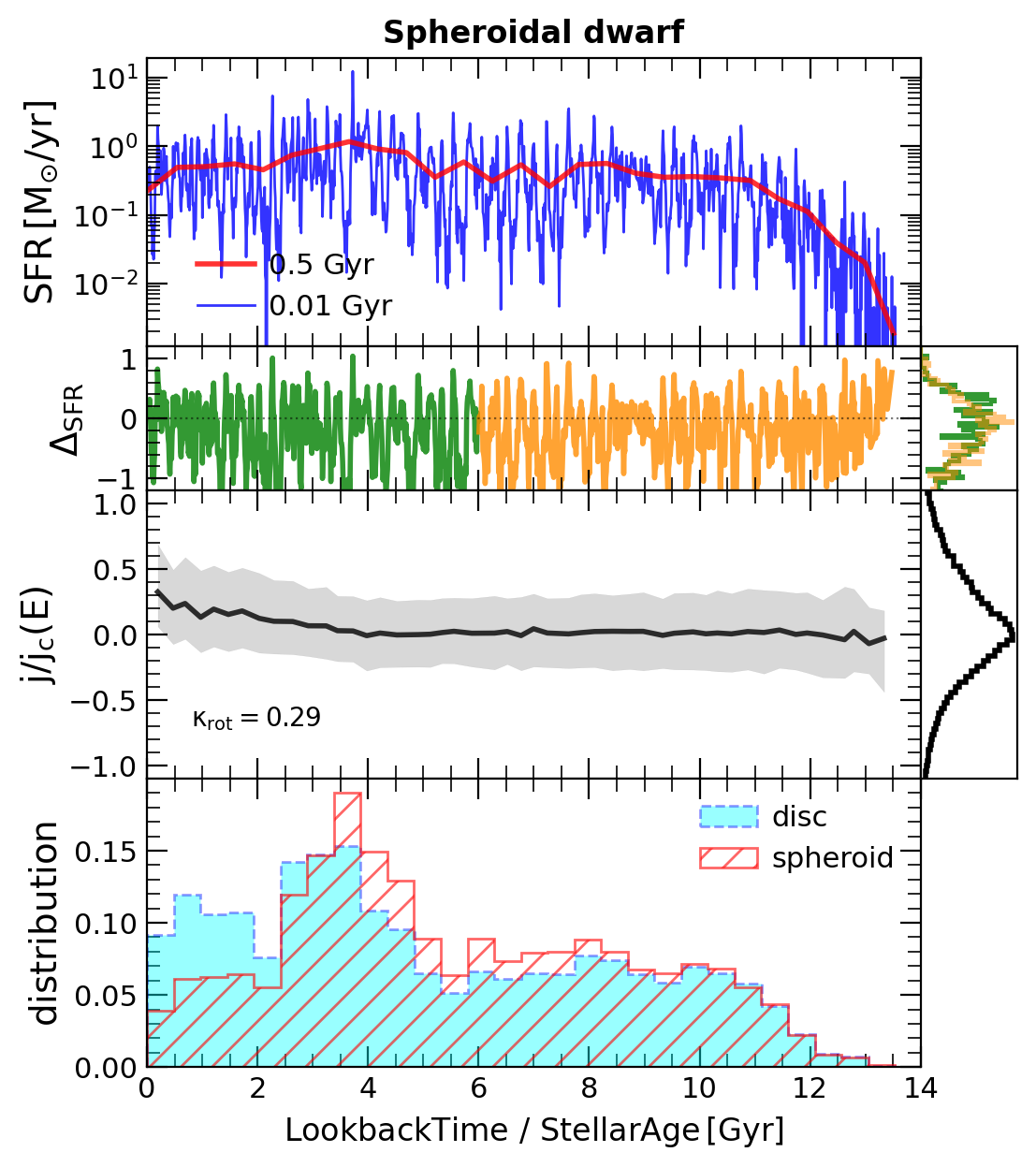}
	\includegraphics[width=0.325\textwidth]{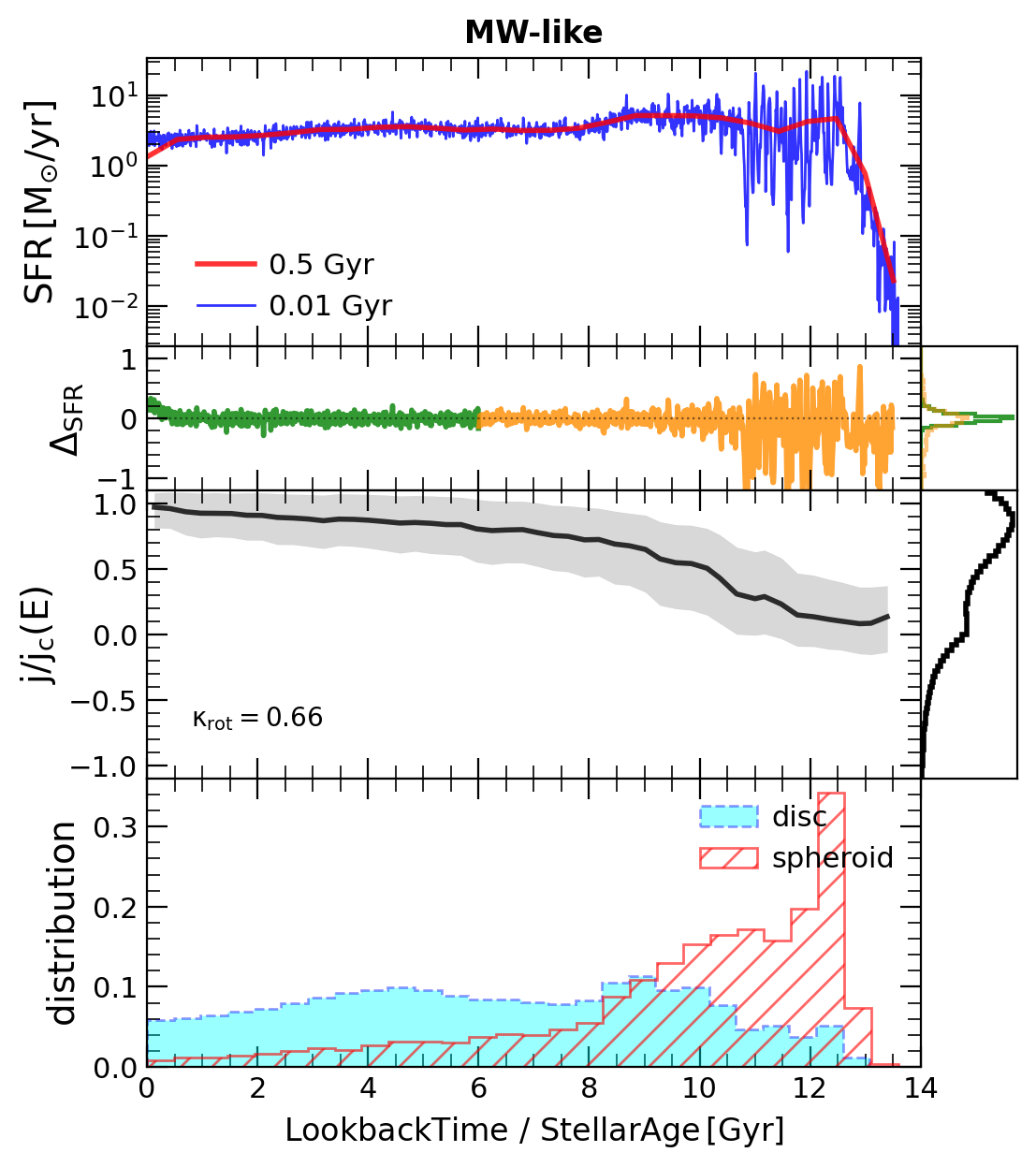}
	\caption{Several evolutionary properties for the three example galaxies in Fig~\ref{fig:krot_Mstar}, disky dwarf (left), spheroidal dwarf (center), and MW-like (right). First row: star-formation rate (SFR) as a function of lookback time averaged in two time-scales, 500 Myr in red, and 10 Myr in blue. Second row: the logarithmic difference between the SFR curves averaged in 10 Myr with respect to the 500 Myr one, $\rm{\Delta_{SFR}}$. The distribution of $\rm{\Delta_{SFR}}$ above and below 6 Gyr are shown in the right box with green and orange colors respectively. Third row: median of the stellar circularities ($\epsilon = j/j_c (E)$), and their distribution in the vertical histogram. Fourth row: distribution of stellar particles at present day associated with the disk ($\epsilon > 0.5$ in blue) and spheroid ($\epsilon < 0.5$ in red) components in each galaxy. For the MW-like galaxy, burstiness subsides $\sim 10$ Gyr ago and a stellar disk begins to grow with circularities $j/j_c \sim 1$. For the dwarfs, bustiness continues throughout their evolution, with the spheroid-dominated one showing larger bursts (wider $\rm{\Delta_{SFR}}$ variations).}
	\label{fig:examples}
\end{figure*}

\section{The Origin of the morphological transition in the FIRE simulated low-mass galaxies}
\label{sec:main_results}

\subsection{Halo spin}
\label{ssec:spin_halo}

To quantify the influence of dark matter halo dynamics on the stellar component in Figure~\ref{fig:lambda} we present the dimensionless halo spin parameter ($\lambda$) as a function of stellar mass for all simulated central galaxies. We have calculated the $\lambda$ halo spin following \citet{Bullock2001},
\begin{equation}
\lambda = \frac{J}{\sqrt{2} M_{\rm vir} V_{\rm vir} r_{\rm vir} } \ ,
\label{eq:lambda}	
\end{equation}
where $J$ is the total angular momentum of dark matter particles within the virial radius ($r_{\rm vir}$). The distribution of the halo spin parameter of all central galaxies in FIREbox (cyan circles) presents a median value of $\lambda_{\rm all}=0.031^{+0.014}_{-0.010}$, independent of $M_\star$. The \textit{m11} zoom-in runs (gray diamonds) show a median of $\lambda_{\rm{m11's}}=0.041^{+0.021}_{-0.018}$  which is both consistent with FIREbox and \citet{Bullock2001}. As expected from $\Lambda$CDM, we confirm that the halo spin does not vary noticeably over the stellar mass range probed in our sample.\\

It has been shown that the halo spin parameter plays a negligible role in the morphology of galaxies with masses comparable to the MW using the FIRE-2 model \citep{GarrisonKimmel2018}, a trend reported also in other simulations \citep[e.g., ][]{Sales2012}. However, for low mass galaxies, halo spin can play a more significant role \citep{RodriguezGomez2017}. Here, we explore the link between $\lambda$ and morphology for galaxies in FIREbox with $M_\star = 10^{7.5} - 10^{10} ~ \rm{M_{\odot}}$. Yellow and red circles in Fig.~\ref{fig:lambda} show the spin parameter sub-populations with excess rotational support (high-$\kappa_{\rm rot}$) and a deficit of rotational support (low-$\kappa_{\rm rot}$), selected as objects above of 90th percentile and below 10th percentile, in the $\kappa_{\rm rot} - M_\star$ relation, respectively.\\

We find a weak (but statistically significant) trend of morphology with halo spin for our low mass dwarfs. The median $\lambda$ values for high (yellow) and low (red) $\kappa_{\rm rot}$ samples are indicated on the vertical histogram, and suggest that dwarfs with more rotation support in their stellar components tend to occupy halos with larger spins. The median spin of high-$\kappa_{\rm rot}$ dwarfs is $\lambda_{\rm{high}-\kappa_{\rm rot}}=0.051^{+0.019}_{-0.014}$ (uncertainties corresponding to the 25th$-$75th percentiles), while the median for low-$\kappa_{\rm rot}$ dwarfs is $\lambda_{\rm{low}-\kappa_{\rm rot}}=0.025^{+0.008}_{-0.009}$. Despite this median trend \citep[which agrees with findings from a different galaxy formation model presented in][]{RodriguezGomez2017}, there is substantial overlap in the spin of dwarfs with most and least rotation support, highlighting that while halo spin does play some role in defining the rotation support in low-mass galaxies, it is not sufficient to predict the final morphology of the simulated dwarfs.

\begin{figure*}
	\centering	
	\includegraphics[width=0.485\textwidth] {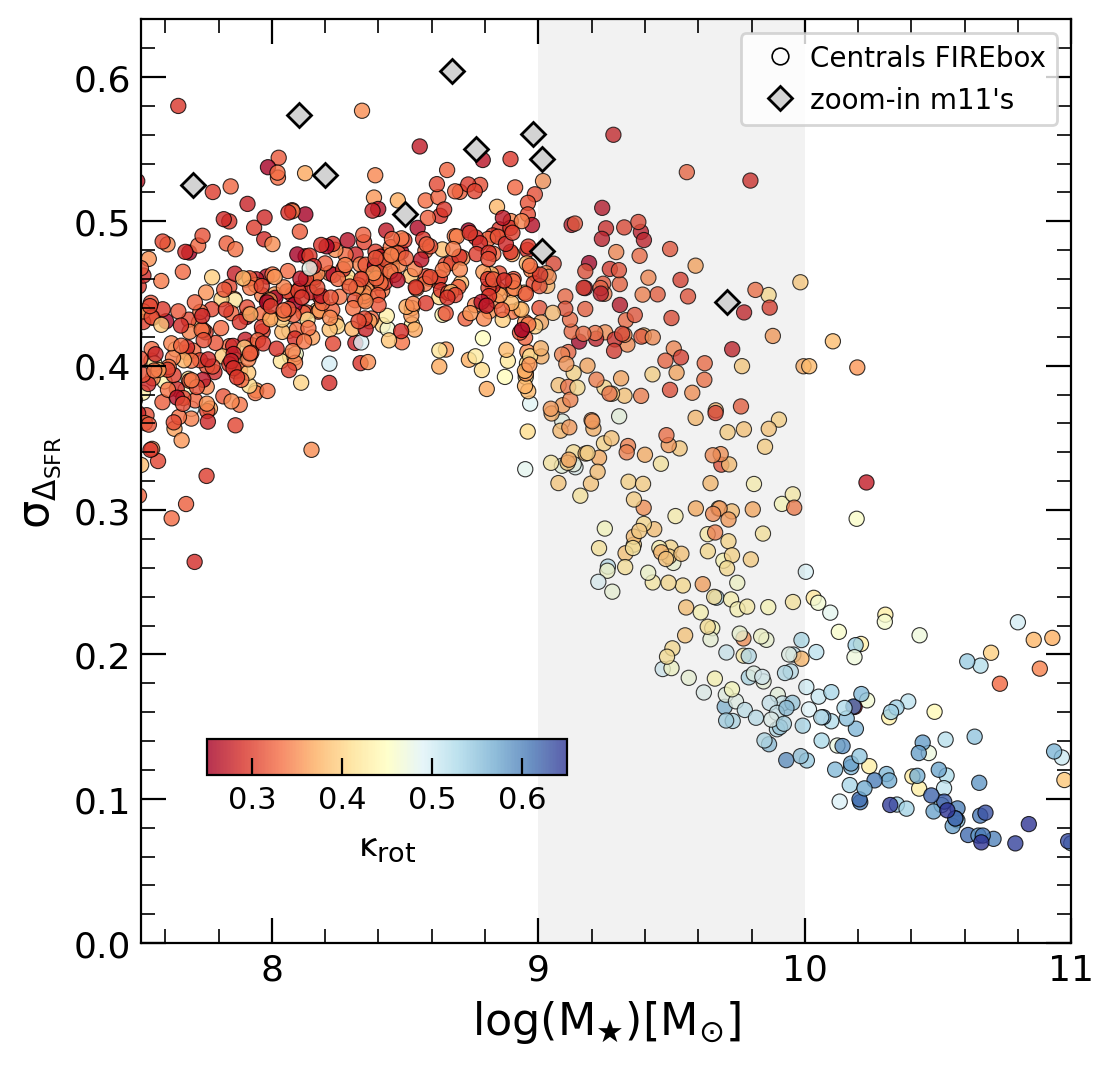}
	\includegraphics[width=0.49\textwidth] {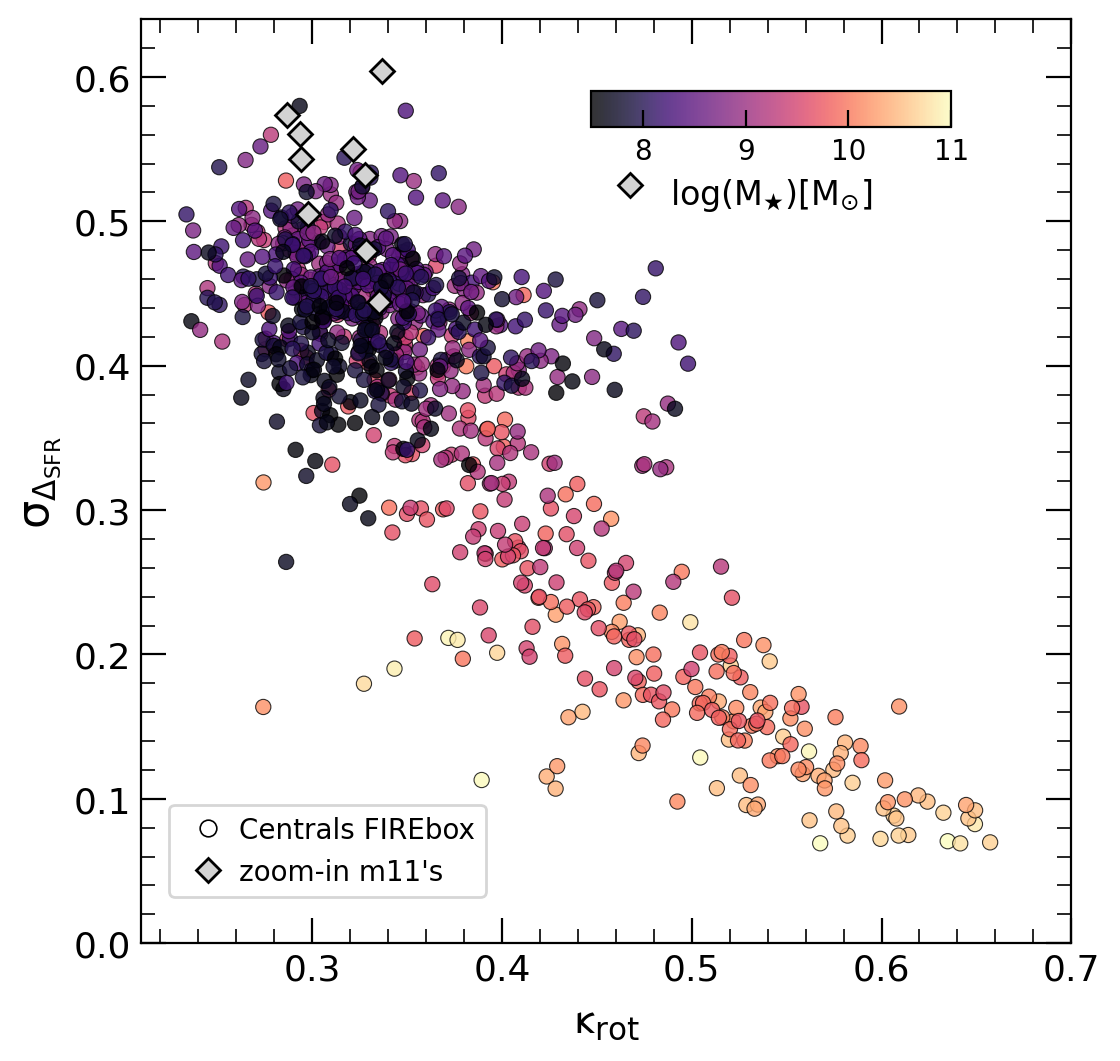}
	\caption{{\it Left:} burstiness as a function of stellar mass, where burstiness ($\sigma_{\Delta_{\rm SFR}}$) is quantified through the dispersion in the $\Delta_{\rm SFR}$ values averaged over the last 6 Gyr. Circles show all central galaxies in FIREbox color codded according to their morphology ($\kappa_{\rm rot}$), blue are disk-dominated, red are spheroid-dominated. Zoom-in m11's FIRE-2 runs are shown as before with grey diamonds, displaying a slightly higher burstiness than FIREbox objects (see text). In FIREbox, burstiness shows a clear decline in the morphology ``transition region'' (gray shaded), after which disk-dominated galaxies become common for $M_\star \geq 10^{10} ~ \rm{M_{\odot}}$. At fixed stellar mass, galaxies with larger $\sigma_{\Delta_{\rm SFR}}$ are more dispersion-dominated (reddish colors) than those with lower burstiness, which tend to have more rotational support. {\it Right:} burstiness as a function of morphology in our sample, confirming a strong correlation between the two variables. Symbols are color coded according to their stellar mass.}
	\label{fig:Delta_SFR}
\end{figure*}

\subsection{Burstiness and the depth of the gravitational potential}
\label{ssec:burstiness}

In the case of MW-like galaxies simulated with the FIRE model, \citet{Yu2021, Yu2023} show that the star formation rate evolves with time, transforming from ``bursty'' at early redshifts --coincidental with the assembly of the spheroidal and thick-disk component -- to a more stable, steady star formation rate at late times that promotes the formation and settling of the thin disk. We show in Fig.~\ref{fig:examples} the star formation rate evolution (top panel) for the three FIREbox galaxies presented in the right panel of Fig.~\ref{fig:krot_Mstar}. The right-most column corresponds to the MW-like analogue and shows a similar behavior as found in \citet{Yu2023} using FIRE-2 zoom-in runs.\\

Burstiness can be quantified by comparing the fluctuations of the star formation rate in two different time scales. Following \citet{FloresVelazquez2021}, we show the SFR averaged over short timescales (10 Myr, blue curve) along with that averaged over longer time scales ($500$ Myr, red curve). For the specific case of the MW-mass object (left column), we see that the SFR is characterized by rapid fluctuations before a lookback time $\sim 10$ Gyr, followed by a quiet-down phase onwards where the short-term averaged SFR coincides nicely with the SFR averaged over longer timescales. \\

We define $\Delta_{\rm SFR}$ to quantify ``burstiness'' as a function of time, which is defined here as the logarithmic difference between the short-term averaged SFR (10 Myr) and the long-term averaged SFR (500 Myr). In the second panel from the top, we show that this parameter nicely captures the behavior described above for the MW-mass object, displaying large values for a lookback time $\geq 10$ Gyr, but evolving smoothly afterwards. During this later, more quieter SFR phase, the circularity of the stars in the galaxy steadily increases (third panel top to bottom), signaling the build up of the thin disk component during the smooth SFR phase. Here, circularity is calculated as the ratio between the z-component of the angular momentum of each star in a system rotated so that the total angular momentum of the stars points along the $z$-direction, compared to that of a circular orbit with the same energy \citep[$\epsilon=j/j_{\rm circ}(E)$, ][]{Abadi2003}. The bottom panel shows the age distribution of the stars selected to belong to the disk ($\epsilon>0.5$, cyan) or to the spheroid ($\epsilon<0.5$, red), confirming the early formation of the stars that are today in the non-rotating component.\\

We extend this analysis to the regime of low mass galaxies, where the left and middle panels of Fig.~\ref{fig:examples} correspond to the disky and spheroidal dwarfs illustrated in Fig.~\ref{fig:krot_Mstar}. Contrary to MW-like galaxies, both dwarfs seem to maintain a substantial level of burstiness throughout their entire evolution. In the case of the disky dwarf (left), the mean circularity of the stars only recently ($\sim 2$ Gyr ago) starts to increase towards $\epsilon \sim 1$, indicating a much younger (and less dominant) disk than in the case of MW-like galaxies. There is also a subtle hint suggesting that the spheroidal dwarf has a larger degree of burstiness in its star formation rate overall, as quantified by a larger dispersion in $\Delta_{SFR}$ values during its entire time evolution.\\

Individual inspection of these example galaxies suggest that the root mean square (r.m.s.) dispersion of the distribution of $\Delta_{SFR}$, or $\sigma_{\Delta_{SFR}}$, can be a helpful indicator of burstiness if integrated over a given time. Inspired by the results on the scale of the MW-like galaxies, we chose to analyze the late-time burstiness as the r.m.s. dispersion of $\Delta_{SFR}$ in the last 6 Gyr (range highlighted in green on the second row of Figure~\ref{fig:examples}). We note that the results presented below do not change qualitative if we instead vary this $6$ Gyr cutoff by $\pm 2$ Gyr. We show on the left panel in Fig.~\ref{fig:Delta_SFR} burstiness (quantified through $\sigma_{\Delta_{SFR}}$) as a function of stellar mass, or the $\sigma_{\Delta_{SFR}} - M_{\star}$ relation. Simulated galaxies are color coded according to their morphology, with disk-dominated objects in blue and spheroid-dominated ones in red (see color bar).\\

\begin{figure*}
	\centering
	\includegraphics[width=0.485\textwidth] {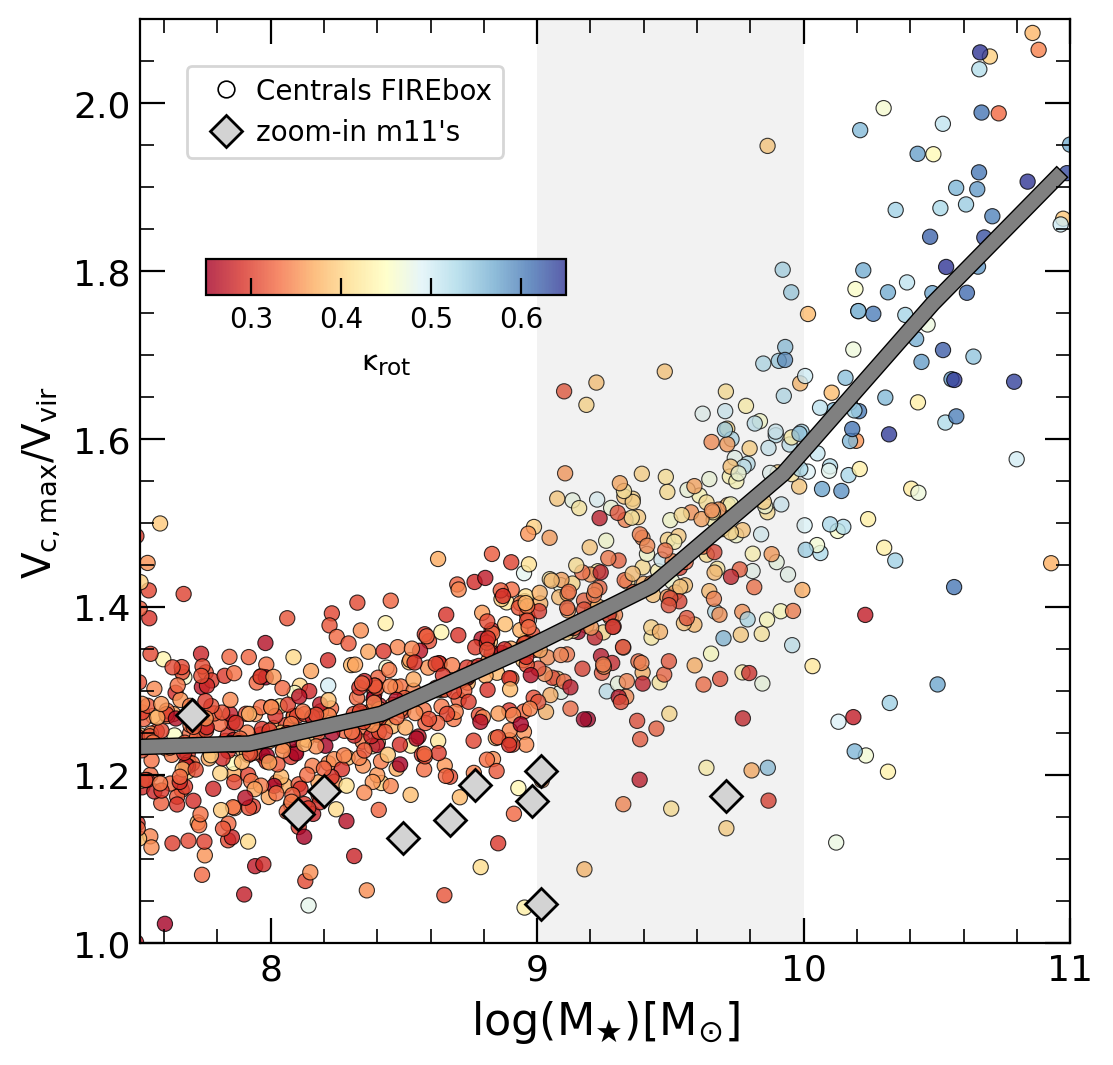}
	\includegraphics[width=0.49\textwidth] {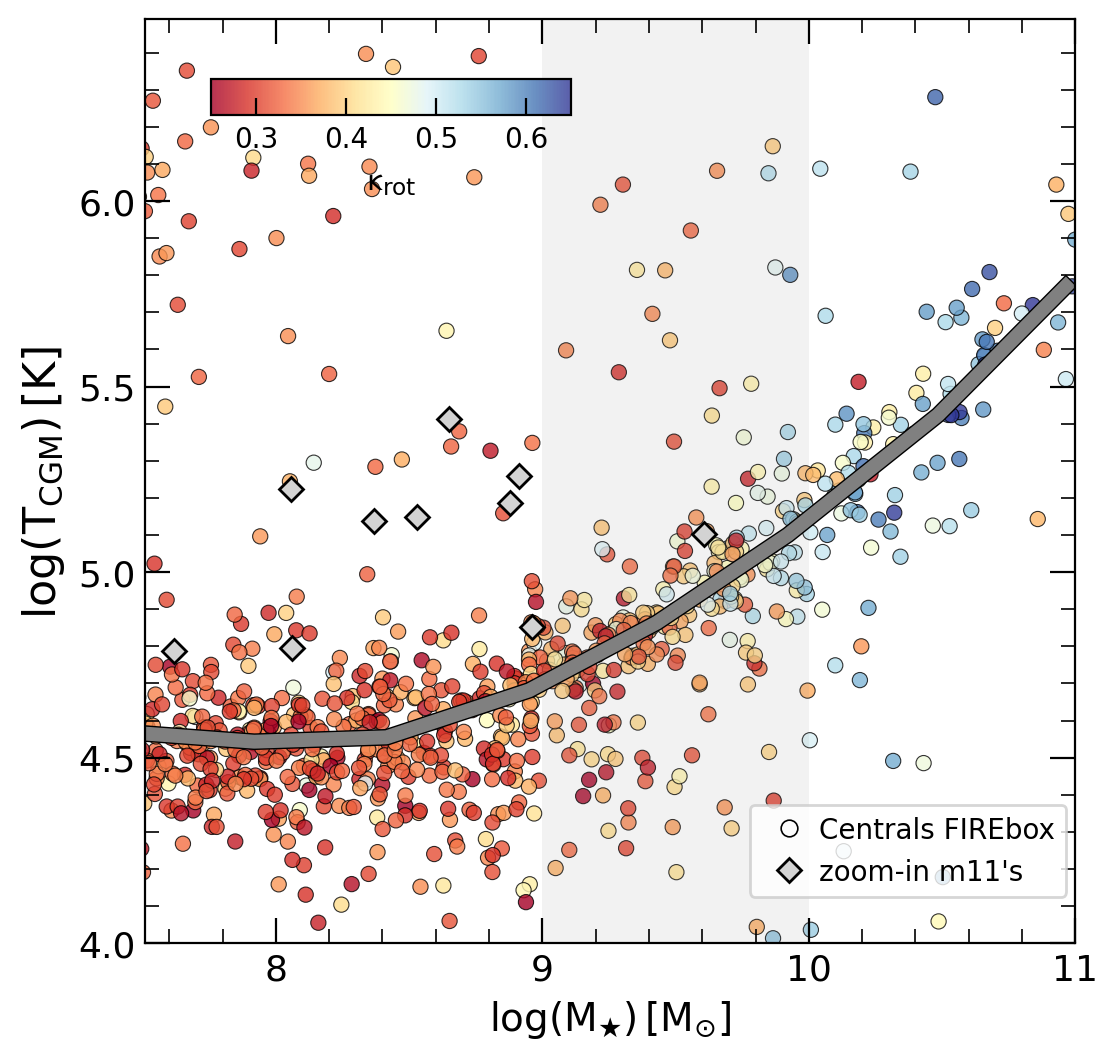}	
	\caption{{\textit{Left:}} Depth of the potential well, here quantified as the ratio between maximum rotation velocity ($\rm{V_{max}}$) and virial velocity of the halo ($\rm{V_{vir}}$), as a function of stellar mass for all central galaxies in FIREbox. Color coding is the same as in Fig.~\ref{fig:Delta_SFR} (see text for discussion of m11's runs). \textit{Right:} Average temperature of the CGM for each halo, calculated as the average temperature of all gas particles in the halo excluding the central regions ($0.3 < r/r_{\rm vir} < 1$). The gravitational potential depth and the temperature of the CGM steadily increases with stellar mass for $M_{\star} > 10^8 ~ \rm{M_\odot}$, facilitating the formation of disks for $M_\star \geq 10^{10} ~ \rm{M_{\odot}}$. }
	\label{fig:VmaxVvir}
\end{figure*}

There is a very clear link between burstiness and morphology in our sample. Two factors are important to appreciate in the left panel of Fig.~\ref{fig:Delta_SFR}. First, the trend of burstiness with stellar mass: $\sigma_{\Delta_{SFR}}$ peaks at values $0.4$ - $0.5$ for stellar masses in the massive dwarf end, $M_\star \sim 10^8 \rm - 10^9~\rm{M_{\odot}}$, and systematically plunge downwards in the morphology ``transition'' region (gray shaded region) reaching $\sigma_{\Delta_{SFR}} \sim 0.1$ at stellar masses comparable to the MW. Second, the scatter in burstiness at fixed stellar mass: galaxies with higher $\sigma_{\Delta_{SFR}}$ show more spheroidal-like components than low $\sigma_{\Delta_{SFR}}$ counterparts with similar stellar content. This effect is maximally visible in the morphology transition region (gray shaded), but is also clear for more massive, MW-like objects. Below $M_\star \sim 10^9 ~ \rm{M_{\odot}}$ the trend disappears since the spread in morphologies is insufficient to display the effect. The strong relation between burstiness and morphology in our sample is further illustrated on the right panel of Fig.~\ref{fig:Delta_SFR}, where we show that $\sigma_{\Delta_{SFR}}$ is smoothly declining as morphologies become more disk-dominated. The overall trend of $\sigma_{\Delta_{SFR}}$ with $M_\star$ and its scatter at fixed stellar mass suggests that the FIRE-2 model predicts too bursty star formation histories to form and maintain rotationally supported stellar disks at low stellar masses. These results also nicely confirm the findings presented in \citet{Yu2023} based on FIRE-2 zoom-in runs of MW-like galaxies, and extend their validity over a wider range of masses, $10^9 < M_\star/\rm{M_{\odot}} < 10^{11}$.\\

Gray diamonds in Fig.~\ref{fig:Delta_SFR} suggest that the higher resolution zoom-in m11 runs roughly agree with results from FIREbox, although zoom-ins are located systematically at higher $\sigma_{\Delta_{SFR}}$. We have explicitly checked that re-computing $\sigma_{\Delta_{SFR}}$ from m11 runs using only an adjusted fraction of the particles to match the sampling in star formation history from FIREbox objects does not resolve the bias. Our definition of $\sigma_{\Delta_{SFR}}$ is relatively robust to the number of particles used to measure it, as far as more than $\sim 1000$ stellar particles are used. Instead, we attribute this slightly higher $\sigma_{\Delta_{SFR}}$ in zoom-in runs to the ability of high resolution simulations to better represent stochasticity associated to star formation than lower resolution runs.\\

Several factors coincide in time with the onset of a less bursty star formation history in objects with MW-like mass. Among them, a deepening of the gravitational potential of the halo \citep{Hopkins2023} along with the establishment of a ``virialized'' or hot CGM surrounding the central galaxy \citep{Stern2021, Hafen2022, Yu2023,Gurvich2023} have been reported to aid the settling of disks. We explore this in Fig.~\ref{fig:VmaxVvir}. In the left panel we use the ratio between the maximum circular velocity of the halo, $V_{\rm c,max}$, to the virial velocity, ($V_{\rm vir} = \sqrt{G M_{\rm vir}/r_{\rm vir}}$) to quantify the depth of the gravitational potential. On the right panel we show the average temperature of the CGM gas, defined here as all gas associated to the halo in the radial range $0.3 < r/r_{\rm vir} < 1$. Symbols are color coded according to morphology ($\kappa_{\rm rot}$) and a thick gray line shows the median of the sample at a given $M_\star$.\\

We find that both, the velocity ratio and the temperature of the CGM monotonically increases with stellar mass, supporting a scenario where disks preferentially arise in halos that are dense and can support a hot circumgalactic medium. Note that the relation found between a hot CGM and the formation of disks differs from the classical picture where hot CGM is associated with red elliptical galaxies \citep[e.g.,][]{Dekel2006}. Instead a hot CGM promotes disk formation by supplying, through cooling, coherently aligned gas to feed the central gas disk \citep{Sales2012, Ubler2014}. Low mass halos hosting dwarf galaxies have virial temperatures $T < 10^5\; \rm K$, making a stable hot CGM rather rare, helping explain the scarcity of disks in that regime \citep{Zeng2024}.\\

Of the factors explored in our FIREbox sample: burstiness (Fig.~\ref{fig:Delta_SFR}, potential depth and CGM temperature (Fig.~\ref{fig:VmaxVvir}), burstiness shows a better and more direct correlation with individual morphologies for $M_\star > 10^9 ~ M_{\odot}$, while potential depth and CGM temperature supports more of an average trend with stellar mass at least $> 10^8 ~ \rm{M_{\odot}}$. As in Fig.~\ref{fig:Delta_SFR}, \textit{m11} zoom-in runs tend to agree with results of FIREbox objects, although with a small bias towards lower gravitational potential in the inner regions (lower $V_{\rm max}/V_{\rm vir}$). This is consistent with a scenario where higher resolution zoom-in runs are able to resolve more of the bursty cycle of star formation leading to the formation of central dark matter cores that are slightly more pronounced  than those formed in the lower-resolution FIREbox dwarfs of similar mass.\\

For completeness, we have explored other ways to quantify the depth of the gravitational potential, obtaining similar results as shown above. For example, one can replace the maximum circular velocity by the circular velocity at a more internal radius, perhaps more connected to the physical scales of the gas and stars in galaxies. For this, we used the circular velocity measured at the fiducial radius $V_{\rm fid}=V_{c}(r_{\rm fid})$, where $r_{\rm fid}=2(V_{\rm max} /70 ~ \rm{km \, s ^{-1}} )$ kpc as introduced in \citet{SantosSantos2020}, with typical values for $r_{\rm fid} \sim 1.3 - 9.5$ kpc for $M_\star \sim 10^8 \rm - 10^{11} ~ M_{\odot}$. The trend using $V_{\rm fid}/V_{\rm max}$ to quantify the potential depth is comparable to using $V_{\rm max}/V_{\rm vir}$ in Fig.~\ref{fig:VmaxVvir}.

\subsection{Gas dynamics}
\label{ssec:gas_ans_young_stars}

The absence of coherent rotational structure in galaxies can result from two different scenarios: $A)$ either stars initially form in a rotationally supported disk that is later disrupted (e.g., by mergers or misaligned gas accretion), or $B)$ stars never form in a rotationally supported configuration to begin with}. We study this in Fig.~\ref{fig:krot_gas}, where we show $\kappa_{\rm rot}$ of the cold gas (pink) as a function of stellar mass. Here, the morphological parameter $\kappa_{\rm rot}$ is calculated as before (Eq.~\ref{eq:krot}), but using cold gas elements within the galaxy radius instead of the stars, where cold refers to elements with temperature $T < 1.5 \times 10^4 ~ K$. The median of the population in FIREbox is shown by the long dashed curve, while shaded regions indicate the 10th$-$90th percentiles in the sample. The trend for the rotational support of the gas mimics closely that of the stellar morphology in simulated galaxies: low mass dwarfs have little rotational support on their gaseous component, which starts to steadily increase in the range $M_\star=10^9 \rm - 10^{10} ~ M_{\odot}$, reaching $\kappa_{\rm rot} \sim 0.9$ for MW-like galaxies. Note that while the trend is similar for gas and stars, stars (blue) shows substantially lower rotation than the gas component.\\

\begin{figure}
	\centering
	\includegraphics[width=\columnwidth] {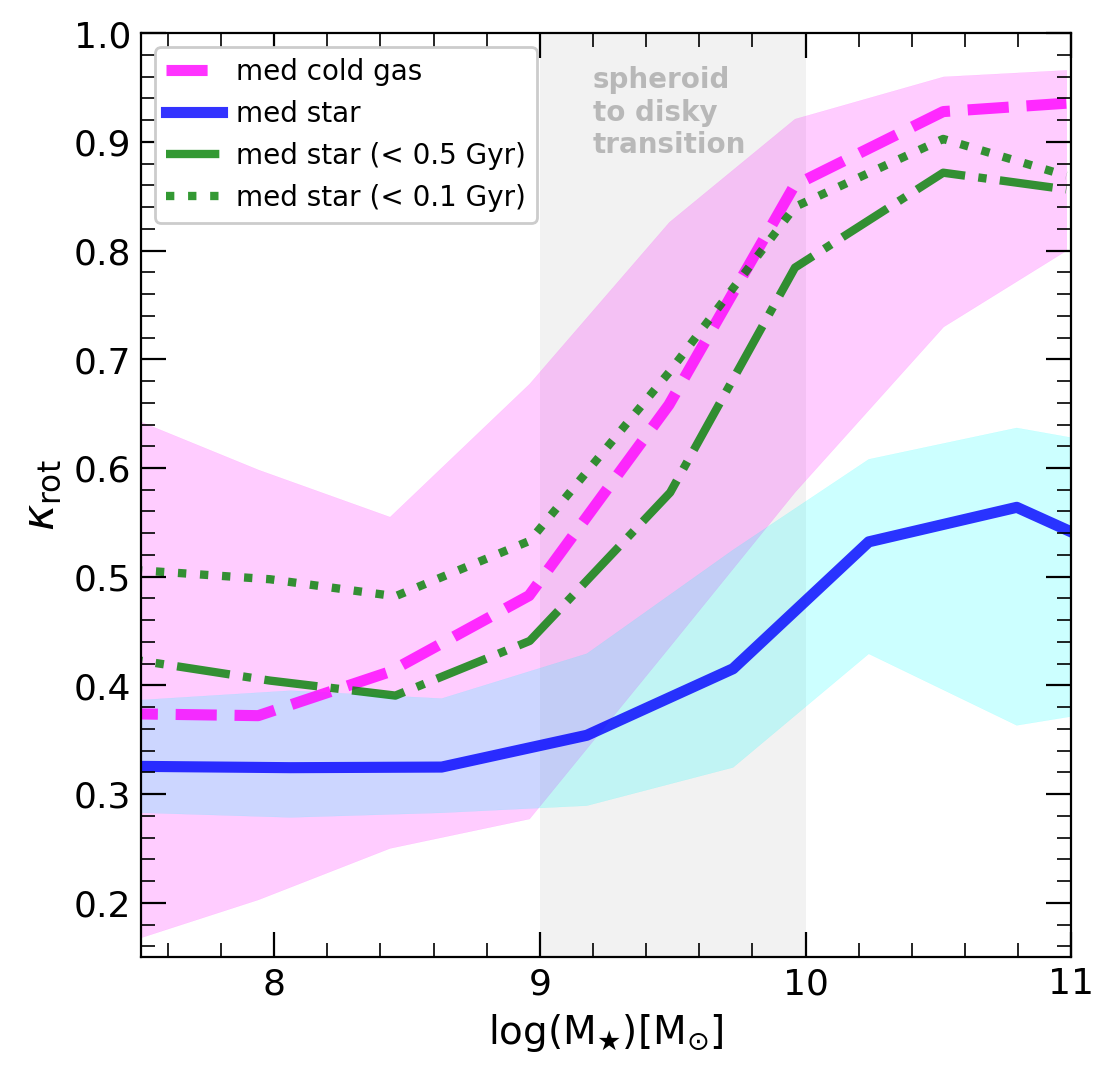}
	\caption{Rotational support of several galaxy components ($\kappa_{\rm rot}$) as a function of stellar mass. Pink dashed line shows the median of the $\kappa_{\rm rot}$ for the gas in all galaxies from FIREbox, along with the shaded region highlighting 10th$-$90th percentiles. Gas shows in general more rotational support than the stars (blue curve), but follows a similar trend: $\kappa_{\rm rot}$ values for the gas increase steeply in the ``transition region'' (gray shaded) reaching almost perfect rotational support for MW-like objects. We show that young stars are born with similar orbits than the cold gas, shown here by the median $\kappa_{\rm rot}$ of stars younger than 500 Myr (dot-dashed) and 100 Myr (dotted green). Low mass galaxies, $M_\star < 10^9~ \rm{M_{\odot}}$, show little rotational support in the stars because even for the gas component dispersion dominates the kinematics, explaining the lack of disks for the simulated dwarfs.} 
	\label{fig:krot_gas}
\end{figure}

Results in Fig.~\ref{fig:krot_gas} confirm the origin of the lack of disks in low mass simulated galaxies: the gas is unable to settle into rotationally supported components from which stars can be born \citep{Kaufman2007}. This is further demonstrated by the green solid and dotted lines, showing the median $\kappa_{\rm rot}$ of stars that are younger than $0.5$ and $0.1$ Gyr, respectively. Unsurprisingly, the morphology of young stellar populations follows that of the cold gas that fuels their formation. In low mass objects with $M_\star < 10^9 ~ \rm{M_{\odot}}$ the absence of gas disks leads to the formation of stellar components that lack rotational support. This rules out scenario $A)$ above, where the lack of disks is explaining by stars born in misaligned disks. Rather, the lack of disks for simulated galaxies with $M_\star< 10^9 ~ \rm{M_{\odot}}$ in FIREbox can be attributed to the turbulent nature of the gas in those systems and directly linked to the coupling of stellar feedback to the interstellar medium in dwarfs \citep{Kaufman2007}.\\

\begin{figure*}
	\centering
	\includegraphics[width=1.0\textwidth]{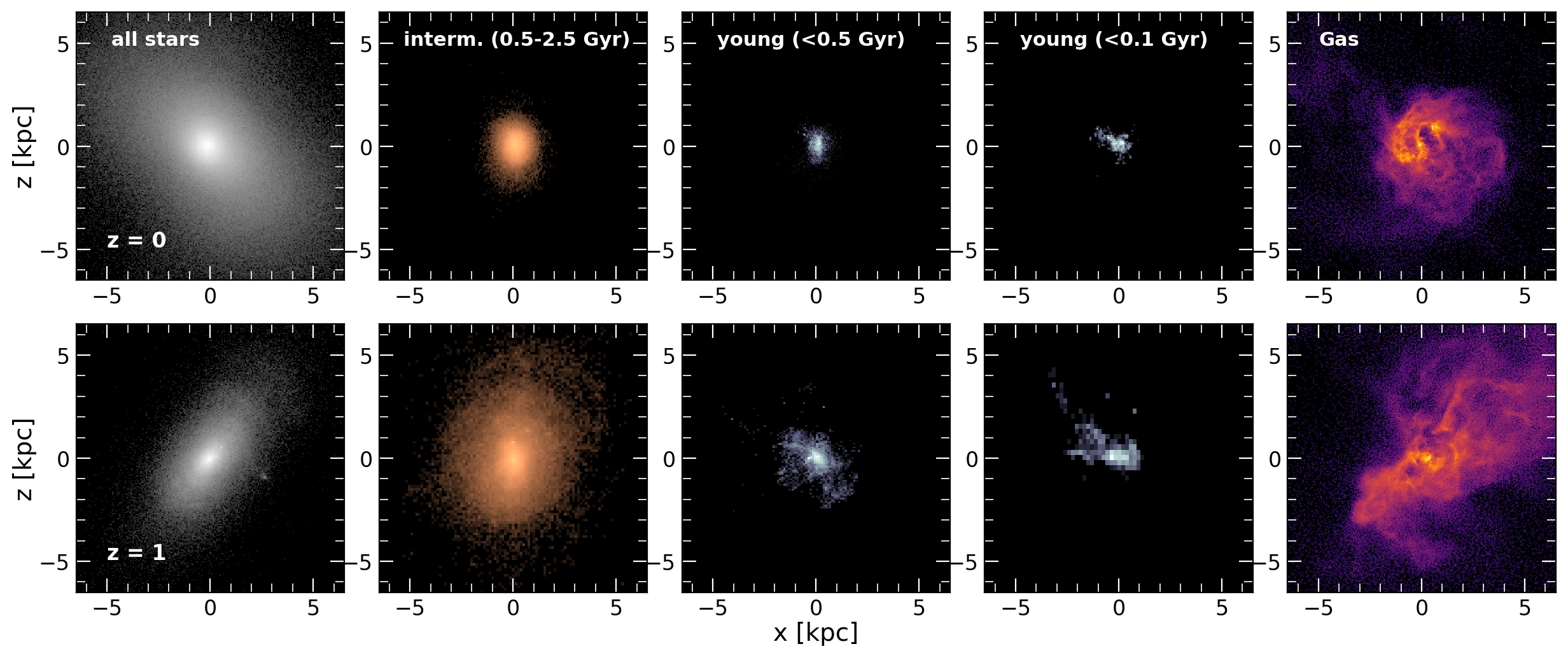}
	\caption{Projections of the stellar (first four columns) and gas (fifth column) components at redshift $z=0$ (top row) and $z=1$ (bottom row) for the \textit{m11d UVB 880} zoom-in run. In the first column, we include all stellar particles, and we divide them by age into intermediate ($0.5-2.5$ Gyr), and young ($<0.5$ or $<0.1$ Gyr) in the subsequent columns. Gas cell distribution is shown in the last column. In each column, we show the edge-on projection, where the system has been rotated with the angular momentum of the displayed stellar particles (or gas cells) pointing along the $z$ direction. In all cases, the structure is irregular and turbulent, with little ordered rotational support and no disk.}
	\label{fig:m11d_star_pops}
\end{figure*}

\begin{figure*}
	\centering
	\includegraphics[width=1.0\textwidth]{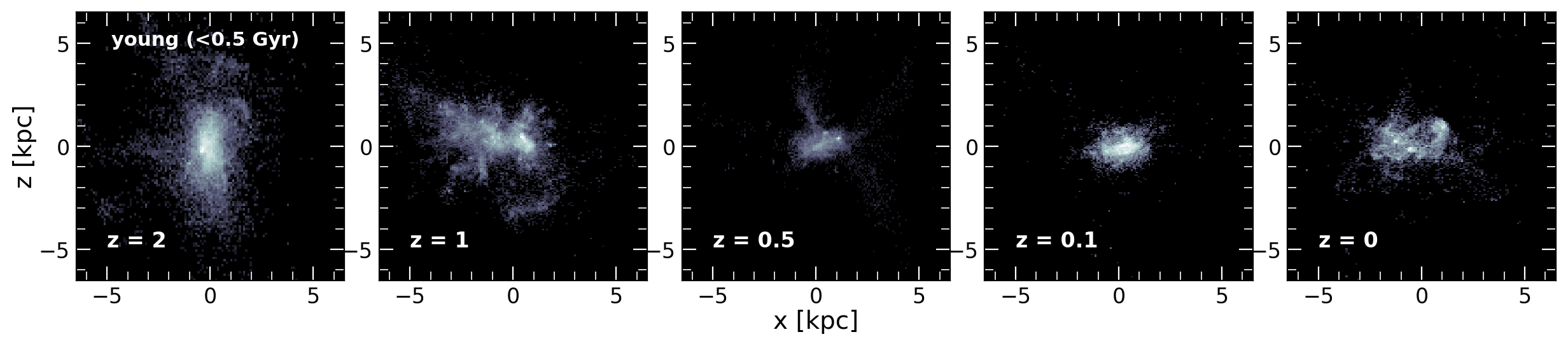}
	\caption{Edge-on projections of the younger stellar population (age $< 0.5$ Gyr) in the \textit{m11q MD 880} zoom-in run at different redshifts (from left to right) $z=[2,1,0.5,0.1,0]$. As the other example shown in Figure~\ref{fig:m11d_star_pops}, this simulation never develops a disk structure throughout time, with their young stars forming with irregular morphologies since $z=2$.}
	\label{fig:m11q_young_stars}
\end{figure*}

Fig.~\ref{fig:m11d_star_pops} illustrates this point further using the higher resolution \textit{m11d UVB 880} zoom-in run (a spheroidal low-mass galaxy with $k_{\rm{rot},z=0} \sim 0.29$). The top row corresponds to the present-day $XZ$ projections of (from left to right) all stars, intermediate, young and very-young stars and gas. Each panel has been individually rotated so that the $z$-component of the angular momentum of each subset of particles is pointing along the $z$-axis (i.e., these correspond to ``edge-on'' views). As is clear from these projections, no axisymmetric disk-like component exists at $z=0$. This statement is also true of an earlier time, $z=1$, shown in the bottom row of the same figure. Moreover, in Fig.~\ref{fig:m11q_young_stars} we select another high-resolution zoom-in run, the \textit{m11q MD 880} ($k_{\rm{rot}, z=0} \sim 0.34$) halo with quietest merger history, and show the edge-on view of the young stars (age $<0.5$ Gyr) at redshifts $z=2, 1, 0.5, 0.1$ and $0$ (left to right). At all times, young stars are born in turbulent structures with no clear sense of coherent rotation.\\

Our results agree well with previous conclusions in \citet{El-Badry2018}, highlighting the turbulent nature of the gas in FIRE zoom-in runs of low mass galaxies, and extend their conclusions to a much larger volume-complete sample from FIREbox. We also demonstrate that this is not a feature of the present-day universe, but rather turbulent structures prevail during most of the cosmological evolution of these dwarfs. On the other hand, for more massive galaxies, once the gravitational potential is able to provide support to contain the bursty cycles, gas morphology becomes dominated by rotation and stellar disks comparable to those of the MW are commonly formed. In other words, the ``morphological transition'' identified in Fig.~\ref{fig:examples}, is a reflection of the mass range where the treatment of the physical processes included in the simulation allows for a transition from turbulent-dominated interstellar medium ($M_\star < 10^9 ~ \rm{M_{\odot}}$) to ordered rotation ($M_\star \geq 10^{10} ~ \rm{M_{\odot}}$).\\

\begin{figure}
	\centering
	\includegraphics[width=\columnwidth] {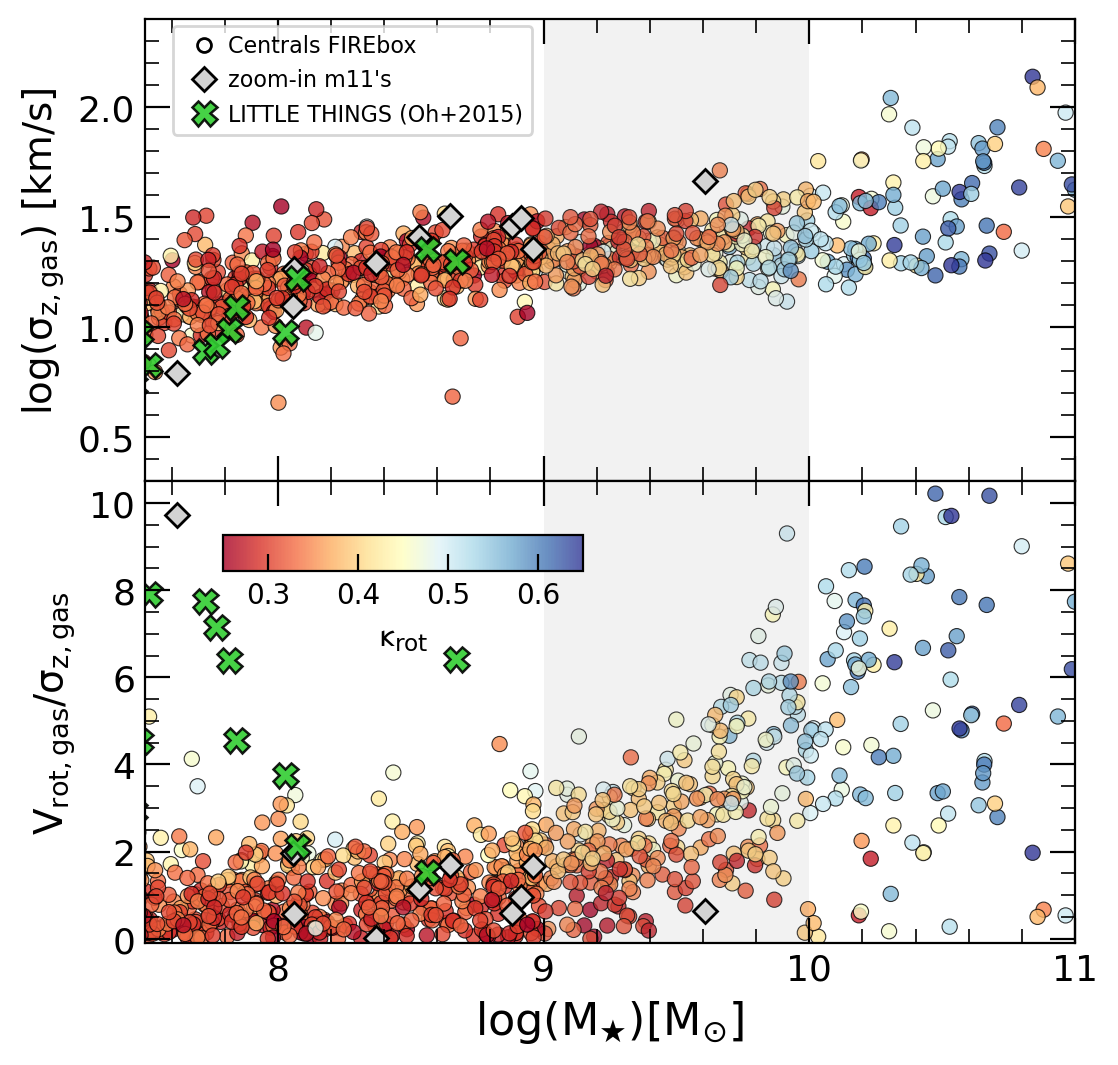}
	\caption{Top: $z$-component of the velocity dispersion in the gas for FIREbox simulated central galaxies color coded by the morphology parameter, $\kappa_{\rm rot}$. The velocity dispersion shows a weak dependence in stellar mass, with $\sigma_{z,\rm gas} \sim 10 \rm - 20$ km/s below $M_\star \sim 10^{10}~ \rm{M_{\odot}}$. Bottom: ratio of gas rotational velocity to velocity dispersion, $\rm{V_{rot} / \sigma}$, as a function of stellar mass. In agreement with the morphological indicator $\kappa_{\rm rot}$, rotational support as measured by $V/\sigma$ in simulated galaxies start growing above 1 in the ``transition region'' with $M_\star=[10^9 \rm - 10^{10}]~$\msun, but remains at very low values for the dwarfs due to the minimum floor in $\sigma_z$. For comparison, we show observational measurements from the rotation curve modelling in 10 LITTLE THINGS galaxies \citep{Oh2015}, indicated with green crosses, which suggest that simulated dwarfs have gas components with larger degree of turbulence than in this sample.}
	\label{fig:V_over_sigma_gas}
\end{figure}

Challenging these results, observations show the formation of rotationally supported disks at lower masses than the transition region in our simulations. Velocity profiles of HI gas  remain double-peaked in observations for velocities where models like FIRE predict a single Gaussian-like distribution \citep{El-Badry2018b}. Similarly, the stellar shape distributions suggest an excess of spheroid-like dwarfs in FIREbox \citep{Klein2025}. We investigate this in our sample in Fig.~\ref{fig:V_over_sigma_gas}, where we show the gas velocity dispersion as a function of stellar mass (top panel) along with the rotational velocity to dispersion ratio (bottom panel). Note that careful comparisons to observational data in the Local Group indicates that the FIRE model reproduces nicely the observed gas dynamics in galaxies above the morphology transition region, or $M_\star \geq 10^{10}$\msun\; \citep{McCluskey2024,McCluskey2025}.\\

The top panel of Fig.~\ref{fig:V_over_sigma_gas} shows that below such scale the velocity dispersion varies only weakly with stellar mass, with characteristic values $\sigma_{z,\rm gas} \sim 10 \rm - 20 ~ \rm{km/s}$ for simulated $M_\star < 10^{10} ~ \rm{M_{\odot}}$ objects. For comparison, we include the average velocity dispersion calculated by the modeling of the rotation curve in a sample of observed dwarf galaxies from LITTLE THINGS \citep{Oh2015}, indicated with green crosses\footnote{For the observations, we have calculated the average $\sigma$ (azimuthally averaged HI velocity dispersion) from column $b)$ of their reported asymmetric drift corrections for each individual rotation curve model.}. The velocity dispersion in the simulation is in reasonable agreement, although there is evidence of a systematically smaller dispersion in LITTLE THINGS for $M_\star < 10^8$\msun. Note that this is a conservative comparison, as the velocity dispersion of the gas in the simulations is calculated only on the $z$-component after each galaxy has been rotated so that the total angular momentum is aligned with the $z$-axis, and not the projected line-of-sight component resulting from the total 3D dispersion.\\

The bottom panel of Fig.~\ref{fig:V_over_sigma_gas} shows the ratio between the rotation velocity $V_{\rm rot,gas}$ to the predicted velocity dispersion $\sigma_{z, \rm gas}$ in the simulations. Here, we calculate $V_{\rm rot,gas}$ in the rotated framework of each galaxy with their stellar angular momentum pointing along the $z$-direction and assign the rotation velocity to the azimuthal component of the velocity: 

\begin{equation}
V_{\rm rot} = V_\phi = \frac{x v_y - y v_x}{x^2 + y^2} \ ,
\label{eq:v_phi}	
\end{equation}

\noindent
Once we have both radial profiles ($V_{\rm rot}, \sigma_z$) we measure their ratios at the galactic radius ($r_{\rm gal}$). The bottom panel in Fig.~\ref{fig:V_over_sigma_gas} illustrates that only galaxies with $M_\star > 10^9~ \rm{M_{\odot}}$ are massive enough to have $V_{\rm rot,gas} \geq 3\sigma_{z,\rm gas}$, giving raise gradually to the formation of rotationally-supported disks in the morphology transition region $M_\star=10^9 \rm - 10^{10}~ \rm{M_{\odot}}$ and beyond. This introduces a minimum halo mass roughly $M_{\rm vir} \sim 10^{11}~ \rm{M_{\odot}}$ where disks can form. Instead, most of the LITTLE THINGS galaxies (green crosses) show  $V_{\rm rot,gas}/\sigma_{\rm gas} > 3$ despite their low stellar mass (and inferred low dark matter halo). While LITTLE THINGS are not representative of the majority of dwarfs (see discussion below), Fig.~\ref{fig:V_over_sigma_gas} demonstrates the existence of real low mass galaxies with substantially larger rotation support than the ones formed within FIREbox.\\

Besides the flat $\sigma$-$M_\star$ relation below $M_\star > 10^{10} ~ \rm{M_{\odot}}$ highlighted in the top panel of Fig.~\ref{fig:V_over_sigma_gas}, there is an additional contribution to the low rotation velocity in our simulated galaxies: the halo mass - stellar mass relation. Fig.10 in \citet{Feldmann2023} indicates that simulated dwarfs in FIREbox occupy smaller dark matter halos than expected from available abundance matching relations. With lower $M_{200}$ at a given stellar mass, the rotational velocity (or circular velocity) is expected to be lower than galaxies that inhabit a larger mass dark matter halo, contributing partially to the low $V/\sigma$ ratios in the low mass simulated galaxies.\\

It is important to keep in mind that comparisons between simulations and observations of gas and star kinematics in the regime of dwarfs is challenging. Samples like LITTLE THINGS are heavily biased towards finding rotation (since they are selected for kinematical studies and dynamical modeling) and are far from representative of the overall dwarf population. Our results suggest that simulated dwarf galaxies show less rotation than {\it some} observed dwarfs, for example, from the LITTLE THINGS survey. But this should not be interpreted as a general inadequacy on the structure predicted for {\it all} simulated dwarfs. On the contrary, FIRE-2 and FIREbox runs have been shown to reproduce several of the scaling relations of observed in low-mass systems. For example,  \citet{Klein2024} shows that synthetic mock observation of simulated galaxies in FIREbox are in good agreement with the observed mass-size relation. Similarly, the gas kinematics in FIRE-2 dwarf zoom-in runs is also shown to overlap with many of the observed dwarfs when comparing unresolved HI profiles \citep{El-Badry2018b}, including a good agreement with the observed baryonic Tully-Fisher relation. The reader is referred to Sec. 8.2 in \citet{El-Badry2018} and Sec. 4.1 and 4.2 in \citet{El-Badry2018b} for a detailed discussion of biases in observations and simulations that should be kept in mind when comparing.\\

What our work highlights here is that the same galaxy formation model, when applied in different regimes, predicts a very different typical structure for galaxies: very commonly disk-dominated in the scale of MW-mass galaxies, but increasingly dominated by dispersion as stellar mass falls below $M_\star \sim 10^8~ \rm{M_{\odot}}$. This is a robust prediction that agrees well with observations: dwarf galaxies are found to be rounder and the disks turn thicker at the low mass end \citep{Helmi2012,Roychowdhury2013,Simons2015,Johnson2017}. Fig.~\ref{fig:V_over_sigma_gas} only highlights that the universe is able to create {\it at least some} dwarf galaxies that are rotationally supported in the low mass regime where FIREbox tends to form systems with significant dispersion support. How common such disky dwarfs are in the universe is still unclear and therefore it might require of simulations of larger volumes to generate a comparable system within the FIRE model. A more clear assessment of the severity of disk scarcity in the regime of dwarfs and the exact scale (stellar mass) where this becomes a problem can only be achieved once kinematical studies of volume-complete samples of low mass galaxies become available.\\

Recently, \citet{Celiz2025} also reported a mass - morphology relation (see Fig.~\ref{fig:krot_Mstar}) in the TNG50 simulation, but with a ``transition region'' occurring at lower stellar masses, $10^8 < M_\star/\rm{M_\odot} < 10^9$. In the low-mass end of that range, \citet{Zeng2024} finds that TNG50 dwarfs form their stars in disks, but later episodes of star formation occur in gas disks that are misaligned with previous events, leading to mostly predictions of dispersion-dominated objects for their low mass dwarfs. While the mass - morphology trend in TNG50 is similar to the one found in our sample, the exact nature of the lack of rotationally supported disks in dwarfs seems different than in FIREbox. \citet{Celiz2025} find that low mass galaxies form the majority of their stars in a central unresolved clump, explaining their lack of stars with ordered rotational support. The authors mention that the central clumps are connected to shortcomings of the wind model and the assumption of a hydrodynamically decoupled wind. In contrast, FIREbox includes a completely different treatment of winds where deposition is local and is able to generate self-consistent outflows and a multi-phase interstellar medium. We have also shown that in FIRE, dispersion-dominated dwarfs tend to form their stars not in misaligned disks that cancel their angular momentum, but directly from dispersionally-supported gas.\\

In the case of FIREbox, other factors might come into play to explain the increase of dispersion support for low mass galaxies. For example, comparison with observations seem to suggest that the timescale between bursts of star formation in the FIRE model could be too short for intermediate mass dwarfs with $M_\star \sim 10^8~ \rm{M_{\odot}}$ \citep{Emami2019}. This would result in more active bursty cycles in low mass galaxies, resulting in a higher amount of energy coupling from stellar feedback into the interstellar medium and, therefore, larger expected degree of gas turbulence, in agreement with our detected trend in FIREbox. Other codes have also reported effects on morphology of dwarf galaxies related to the details of the subgrid physics model. For instance, runs from the MARVEL suite \citep{Munshi2021} also suggest that two different models for supernova energy deposition (blastwave vs. superbubble feedback) would lead to differences in the burstiness expected for low mass galaxies, albeit in their simulations, the feedback-related changes in the bursty cycles appear only for lower mass objects than here, closer to the regime of ultrafaint dwarfs. \citet{Zhang2024} shows using the Stars and MUltiphase Gas in GaLaxiEs ({\sc SMUGGLE}) model that the directionality of the winds assumed can have an impact on the level of burstiness in a $M_\star \sim 10^8~ \rm{M_{\odot}}$  dwarf and also its gas and stellar morphology (in addition to its inner dark matter profile). We conclude that, given the strong relation between star formation cycles, the generated stellar feedback, and which fraction of it effectively couples to the surrounding gas, the theoretical predictions for the morphology of dwarf galaxies in galaxy formation simulations are still uncertain, and heavily impacted by the choices and limitations in the numerical implementation of feedback and star formation.

\begin{figure*}
	\centering
	\includegraphics[width=1.0\textwidth]{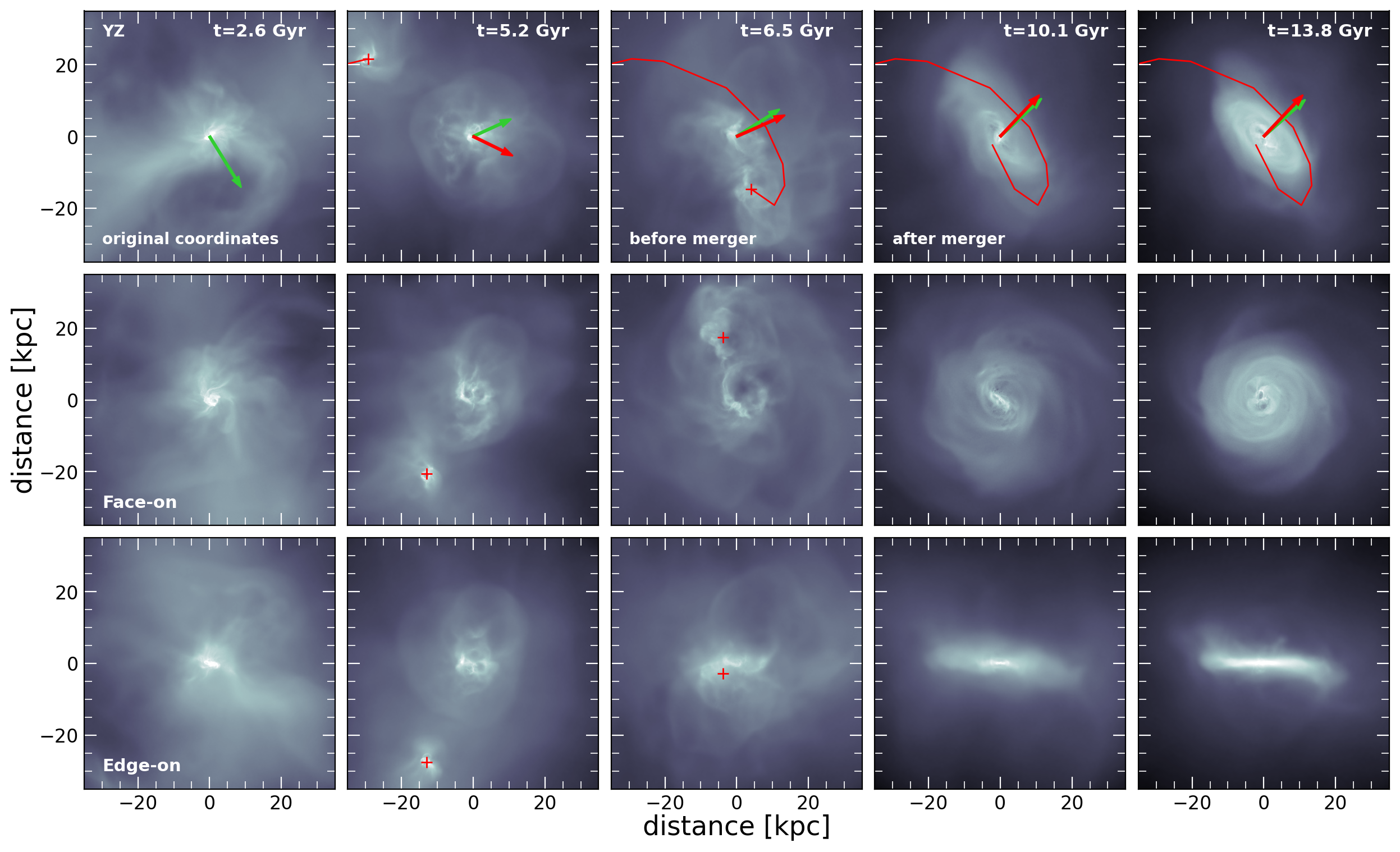}
	\caption{Time evolution for the formation of a disk component in the \textit{m11b} zoom-in run. Top row: each panel shows the gas distribution at different times (columns) in the YZ projection using the original coordinates of the simulation. The light-green arrow indicates the angular momentum vector of the gas (calculated with the gas cells within $\rm{r_{gal}}$), while the red arrow shows the angular momentum vector of the orbit of a satellite responsible for bringing in the gas that forms the rotationally supported disk. We also indicate its orbit with a red solid line for some times before and after the merger. Middle and bottom rows present similar information but rotated to show the face-on and edge-on projections of the system such that the angular momentum of the central galaxy at each time points along the $z$-axis. The position of the satellite is highlighted by the red cross before the merger. The interaction with this gas rich companion mostly explains the formation of a large gas disk in this dwarf, which then fuels the formation of a stellar disk with significant rotation, explaining why \textit{m11b} is a high $\kappa_{\rm rot}$ outlier in the morphology - stellar mass relation (Fig.~\ref{fig:examples}).} 
	\label{fig:m11b_gas}
\end{figure*}

\section{A disk in a low-mass dwarf: the case of \textit{\lowercase{m11b}}} 
\label{sec:m11b}

The left panel in Fig.~\ref{fig:examples} shows that one of the FIRE-2 high resolution zoom-in runs used here, \textit{m11b MD+CR 2100} (hereafter \textit{m11b}), displays a significant amount of rotational support ($\kappa_{\rm rot} \sim 0.47$) that is well above the median $\kappa_{\rm rot} \sim 0.3$ seen in counterparts with similar stellar mass $M_{\star} \sim 4.2 \times 10^{7} ~ \rm{M_\odot}$ and even the $\kappa_{\rm rot}$ for the other m11 zoom-in runs that are more massive. This raises the question as to the nature of the rotational support found in the \textit{m11b} run. Note that a similar version o this \textit{m11b} run, with standard FIRE-2 physics, has been previously studied \citep[see e.g.,][]{Chan2018, El-Badry2018,El-Badry2018b} and is also the main object modified and studied in \citet{Hopkins2023}. The analysis of \textit{m11b} presented here uses a version with additional physics, including metal diffusion and cosmic rays. However, we have checked that the stellar mass and morphology in this particular run are very close to what was obtained with the more traditional flavor of the FIRE-2 model. For example, using the metal-diffusion run, we find $M_{\star}= 5.0 \times 10^{7} ~ \rm{M_{\odot}}$ and $\kappa_{\rm rot}=0.46$, in good agreement with the values in our \textit{m11b MD+CR 2100} run. This confirms that the additional physics included may not fundamentally alter the assembly and resulting morphology in this object, which is the target of study below.\\

We tracked the evolution of the \textit{m11b} host halo and identified a satellite interaction about  $\sim$8.5 Gyr ago ($z \sim 1.2$). The satellite has a stellar mass $M_{\star,\rm sat} \sim 6 \times 10^{6} ~ \rm{M_{\odot}}$ and is gas-rich ($M_{\rm gas,sat} \sim 3 \times 10^{7} ~ \rm{M_{\odot}}$). At this redshift, the stellar mass of the \textit{m11} main host galaxy is $M_{\star,\rm host} \sim 3.3 \times 10^{7} ~ \rm{M_{\odot}}$, meaning that the satellite interaction corresponds to a stellar mass ratio $M_{\star, \rm sat} / M_{\star, \rm host} \sim 0.2$. This interaction seems to have a significant impact on the posterior evolution of the central dwarf. We show this in Fig.~\ref{fig:m11b_gas}, which shows the gas distribution within a box of 60 kpc centered at the main galaxy in the \textit{m11b} run. Different columns correspond to different times throughout its evolution, including before and after the merger at $t_{\rm merger} = 7.5$ Gyr.  The top row corresponds to the distribution of the gas cells, in the $YZ$ projection, by using the original coordinates of the simulation; while the middle and bottom rows show the face-on and edge-on projections, respectively, with the system rotated such that the total angular momentum points along the $z$-axis.\\

The top row illustrates that the interaction occurs such that the orbital momentum of the companion is roughly aligned with the angular momentum of the gas in the central dwarf. To visualize this, we use green arrows in the top row panels to indicate the direction of the total angular momentum of the gas of the central galaxy while the red arrow indicate the direction of the angular momentum of the orbit of this satellite companion. We also use a solid red line to show the projected orbital path followed by the satellite before it merges completely at $t \sim 7.5$ Gyr. For the last 2 columns, we freeze the red line and arrow to those measured at the last snapshot where the satellite is identified before merging. The special alignment of this interaction means that the satellite delivers a substantial amount of gas with large amounts of angular momentum to this object. The middle and bottom rows in Fig.~\ref{fig:m11b_gas} highlight that as a result of this interaction, a very well defined gas disk is formed, which is particularly thin and extended up to a radii $\sim 20$ kpc as a result of the increased angular momentum content originated from the coincidental alignment of the satellite orbit (red crosses highlight the position of the companion in snapshots where it can still be identified by the halo finder). After the disks is formed, star formation occurs in the disk, increasing the $\kappa_{\rm rot}$ of the stars in this galaxy and explaining their excess rotational support compared to other objects in FIREbox or FIRE-2 zoom-in runs. While such interactions are rare in $\Lambda$CDM, this case nicely reminds us that extended rotationally-supported disks may occur even in dwarfs as low mass as $M_\star \sim 10^7~ \rm{M_{\odot}}$ and the frequency of them in observations may provide helpful constraints for the subgrid models in galaxy formation simulations.

\section{Summary}
\label{sec:concl}

In this study, we explore the morphology of galaxies in the stellar mass range $M_\star=[10^{7.5} \rm - 10^{11}]~ \rm{M_{\odot}}$ using the FIREbox simulation \citep{Feldmann2023}, with particular emphasis on the regime of dwarf galaxies. We also use several higher-resolution zoom-in runs from the m11's suite (halo mass $M_{\rm vir} \sim 10^{11} ~ \rm{M_{\odot}}$)  also run with the FIRE-2 model to study individual objects with a higher time-cadence and larger numerical resolution. We characterize morphology using the dynamical criteria $\kappa_{\rm rot}$ which measures the fraction of the kinetic energy in ordered rotation \citep{Sales2012}.\\

Overall, we find that FIREbox is able to reproduce a  variety of morphologies in the galaxy population, ranging from dispersion-dominated spheroidal systems to rotational-supported thin disks. We identify a clear trend  between morphology and stellar mass that supports the idea that low mass galaxies are dispersion-dominated while galaxies comparable to the MW often host dominant stellar disks. In particular, we find that low mass dwarfs with $M_\star < 10^9~ \rm{M_{\odot}}$ are rarely disk-dominated, while more massive galaxies with $M_\star > 10^{10}~ \rm{M_{\odot}}$ are mostly rotationally supported. Therefore, FIREbox predicts a ``transition regime'' from $M_\star=10^9 \rm - 10^{10}~ \rm{M_{\odot}}$, where disks start to arise and become the dominant dynamical component for galaxies of mass comparable to the MW. Some evidence exists in support of a similar transitional mass in observations \citep{Simons2015}.\\

We investigate the physical factors that determine this morphology trend with stellar mass in our simulations. We find that, contrary to galaxies of $\sim$MW-mass, the spin of the halo can play a significant role in the rotational support of the stellar component in low mass galaxies ($M_\star < 10^{10}~ \rm{M_{\odot}}$), in agreement with previous claims using different simulations and baryonic treatment \citep[e.g.][]{RodriguezGomez2017}. However, there is no indication that halo spin plays a role in establishing the morphology transitional mass or the overall trend of morphology with stellar mass.\\

Instead, we find that the burstiness of the star formation history (here characterized by $\sigma_{\Delta_{\rm SFR}}$, Fig.~\ref{fig:Delta_SFR}) shows a strong correlation with both, the morphology-mass relation as well as its scatter at fixed stellar mass. In other words, in FIREbox, galaxies of a given stellar mass show higher rotational support in their stars (high $\kappa_{\rm rot}$) if their star formation histories are less bursty than counterparts showing more violent star formation cycles. In particular, measured as the r.m.s dispersion of the star formation history averaged in 0.1 and 0.5 Gyr, the $\sigma_{\Delta_{\rm SFR}}$ parameter shows a strong decline in the stellar range $M_\star=[10^9 \rm - 10^{10}]~ \rm{M_{\odot}}$, coincident with the scales where morphology globally transitions from purely dispersion dominated to mostly disk-dominated. Additionally, we find that the depth of the potential well along with the presence of a hot CGM accompanies the trend in burstiness, showing a systematic increase with stellar mass that follows the occurrence of disks in the simulation.\\

Rotationally supported stellar disks become increasingly rare at low masses in the simulation due to the predicted dispersion support in the gas component, which is tied to the star formation burstiness. The gas velocity dispersion scales only weakly with stellar mass, with lower-bound values $\sigma_z \sim 10$ km/s for $M_\star=10^8 ~ \rm{M_{\odot}}$ and only $\sim 20$ km/s for masses $M_\star > 10^{10.5}~ \rm{M_{\odot}}$. At the typical rotational velocities in this mass range (average $V_{\rm rot}= 10$ km/s and $>160$ km/s respectively),  this corresponds to a gas rotational support $V/\sigma \sim 0.5$-$1$ for low mass objects (and therefore, mostly dispersion-dominated systems), while galaxies beyond the ``transitional mass" $M_\star > 10^{10}~ \rm{M_{\odot}}$ may display substantially more rotational support $V/\sigma \sim 1$-$9$. Stars are born from the gas and therefore inherit its kinematics, explaining the morphological trend with mass in our sample. Only when galaxies occupy sufficiently massive halos (deep potential wells), gravity is able to contain the effects of the bursty star formation cycles, providing an avenue to stabilize gaseous disks and start growing a stellar component with ordered rotation. Our results agree well with previous studies of the collaboration characterizing galaxy morphology using FIRE-2 zoom-in runs  \citep{Stern2021,Yu2021,Yu2023,Hafen2022,Gurvich2023,Hopkins2023}, and extend the analysis to include a continue range of stellar masses, different environments and assembly histories as sampled by the cosmological volume box in FIREbox.\\

In our simulations, the morphological transition occurs for $M_\star=[10^9 \rm - 10^{10}]~ \rm{M_{\odot}}$, which means that MW-like galaxies are often predicted to host dynamically dominant thin disks, but they become rare for $M_\star < 10^9~ \rm{M_{\odot}}$. A similar mass-morphology trend is inferred from observations \citep{Roychowdhury2013, Simons2015}, although the exact mass scale for the transition is not yet well constrained. The lack of volume-complete observational samples at the faint-end also prevents a proper sampling of the dispersion around the median of the mass-morphology relation and a better understanding of the rotation- and dispersion-dominated extremes. For example, a comparison with LITTLE THINGS suggests that rotationally supported dwarfs with $V/\sigma >3$ do exist for real galaxies with $M_\star \sim 10^8~ \rm{M_{\odot}}$, which are absent in our sample. However, selection effects in observational dynamical studies prevents a thorough evaluation of the degree of the disagreement: LITTLE THINGS objects were selected based on their large rotation. Previous studies have found that the FIRE-2 model may overpredict the gas dispersion support in the low mass end \citep{El-Badry2018,El-Badry2018b}. In our case, we find that in addition to a slightly higher velocity dispersion in FIREbox compared to low-mass LITTLE THINGS galaxies, the low rotation velocity (itself associated to their low mass dark matter halos) also contributes to the low $V/\sigma \leq 1$. Models where star formation proceeds less efficiently and dwarfs occupy larger dark matter halos at fixed $M_\star$ --as perhaps predicted by current abundance matching models-- may help form some more rotationally-supported disks in the regime of dwarfs.\\

Recently, studies using the TNG50 simulations also reported a similar morphology-stellar mass trend, albeit their transition region occurs at lower masses, closer to $M_\odot \sim 10^8~ \rm{M_{\odot}}$ \citep{Zeng2024,Celiz2025}. TNG50 and FIRE are very different models, and the origin of the lack of disks in TNG50 seems unrelated to the one found here for FIREbox. Other simulations have also reported substantial changes in morphology or bustiness in the dwarfs regime triggered by variations in the treatment of feedback \citep[e.g., ][]{Zhang2024,Azartash-Namin2024}. While most of these models have successfully reproduced the morphology of galaxies at the scale of the MW, the extrapolation of the subgrid models to the regime of dwarfs highlights larger uncertainties (or disagreement) in this regime. The morphology of low-mass galaxies may therefore provide a more refined test of our galaxy formation models, extending beyond the successes already achieved in understanding the formation of MW-mass disks.

\section*{Acknowledgments}
The authors would like to thank Matthew Smith, Julio Navarro and Simon White for insightful discussions that helped strengthen and improve the earliest versions of this draft. JAB and LVS are grateful for partial financial support from NSF-CAREER-1945310 and NSF-AST-2107993 grants. Some of the computations were performed using the computer clusters and data storage resources of the HPCC, which were funded by grants from NSF (MRI-2215705, MRI-1429826) and NIH (1S10OD016290-01A1). AW received support from NSF, via CAREER award AST-2045928 and grant AST-2107772. We acknowledge PRACE for awarding us access to MareNostrum at the Barcelona Supercomputing Center (BSC), Spain. This work was supported in part by a grant from the Swiss National Supercomputing Centre (CSCS) under project IDs s697 and s698. We acknowledge access to the Swiss National Supercomputing Centre, Switzerland, under the University of Zurich’s share with the project ID uzh18. CRW thanks the generous contribution of the Cal Poly Pomona Department of Physics and Astronomy ICR and discretionary funds for summer research support.

\section*{Data Availability}

This paper is based on snapshots, halo catalogs, and merger trees from the FIRE zoom-ins \citep{Hopkins2014, Hopkins2018} and FIREbox \citep{Feldmann2023} data. Some public data are available at \href{https://fire.northwestern.edu/}{https://fire.northwestern.edu/}. The FIRE-2 zoom-in simulations are publicly available \citep{Wetzel2023} at \href{http://flathub.flatironinstitute.org/fire}{FlatHUB}. The main properties of the galaxy samples, and other products included in this analysis, may be shared upon request to the corresponding author if no further conflict exists with ongoing projects.



\bibliographystyle{mnras}
\bibliography{biblio.bib} 

\begin{thebibliography}{}
\makeatletter
\relax
\def\mn@urlcharsother{\let\do\@makeother \do\$\do\&\do\#\do\^\do\_\do\%\do\~}
\def\mn@doi{\begingroup\mn@urlcharsother \@ifnextchar [ {\mn@doi@}
  {\mn@doi@[]}}
\def\mn@doi@[#1]#2{\def\@tempa{#1}\ifx\@tempa\@empty \href
  {http://dx.doi.org/#2} {doi:#2}\else \href {http://dx.doi.org/#2} {#1}\fi
  \endgroup}
\def\mn@eprint#1#2{\mn@eprint@#1:#2::\@nil}
\def\mn@eprint@arXiv#1{\href {http://arxiv.org/abs/#1} {{\tt arXiv:#1}}}
\def\mn@eprint@dblp#1{\href {http://dblp.uni-trier.de/rec/bibtex/#1.xml}
  {dblp:#1}}
\def\mn@eprint@#1:#2:#3:#4\@nil{\def\@tempa {#1}\def\@tempb {#2}\def\@tempc
  {#3}\ifx \@tempc \@empty \let \@tempc \@tempb \let \@tempb \@tempa \fi \ifx
  \@tempb \@empty \def\@tempb {arXiv}\fi \@ifundefined
  {mn@eprint@\@tempb}{\@tempb:\@tempc}{\expandafter \expandafter \csname
  mn@eprint@\@tempb\endcsname \expandafter{\@tempc}}}

\bibitem[\protect\citeauthoryear{{Abadi}, {Navarro}, {Steinmetz}  \&
  {Eke}}{{Abadi} et~al.}{2003}]{Abadi2003}
{Abadi} M.~G.,  {Navarro} J.~F.,  {Steinmetz} M.,   {Eke} V.~R.,  2003, \mn@doi
  [\apj] {10.1086/378316}, \href
  {https://ui.adsabs.harvard.edu/abs/2003ApJ...597...21A} {597, 21}

\bibitem[\protect\citeauthoryear{{Agertz}, {Teyssier}  \& {Moore}}{{Agertz}
  et~al.}{2011}]{Agertz2011}
{Agertz} O.,  {Teyssier} R.,   {Moore} B.,  2011, \mn@doi [\mnras]
  {10.1111/j.1365-2966.2010.17530.x}, \href
  {https://ui.adsabs.harvard.edu/abs/2011MNRAS.410.1391A} {410, 1391}

\bibitem[\protect\citeauthoryear{{Azartash-Namin} et~al.,}{{Azartash-Namin}
  et~al.}{2024}]{Azartash-Namin2024}
{Azartash-Namin} B.,  et~al., 2024, \mn@doi [\apj] {10.3847/1538-4357/ad49a5},
  \href {https://ui.adsabs.harvard.edu/abs/2024ApJ...970...40A} {970, 40}

\bibitem[\protect\citeauthoryear{{Baldry} et~al.,}{{Baldry}
  et~al.}{2010}]{Baldry2010}
{Baldry} I.~K.,  et~al., 2010, \mn@doi [\mnras]
  {10.1111/j.1365-2966.2010.16282.x}, \href
  {https://ui.adsabs.harvard.edu/abs/2010MNRAS.404...86B} {404, 86}

\bibitem[\protect\citeauthoryear{{Barbani}, {Pascale}, {Marinacci}, {Sales},
  {Vogelsberger}, {Torrey}  \& {Li}}{{Barbani} et~al.}{2023}]{Barbani2023}
{Barbani} F.,  {Pascale} R.,  {Marinacci} F.,  {Sales} L.~V.,  {Vogelsberger}
  M.,  {Torrey} P.,   {Li} H.,  2023, \mn@doi [\mnras]
  {10.1093/mnras/stad2152}, \href
  {https://ui.adsabs.harvard.edu/abs/2023MNRAS.524.4091B} {524, 4091}

\bibitem[\protect\citeauthoryear{{Behroozi}, {Wechsler}  \& {Wu}}{{Behroozi}
  et~al.}{2013}]{rockstar}
{Behroozi} P.~S.,  {Wechsler} R.~H.,   {Wu} H.-Y.,  2013, \mn@doi [\apj]
  {10.1088/0004-637X/762/2/109}, \href
  {https://ui.adsabs.harvard.edu/abs/2013ApJ...762..109B} {762, 109}

\bibitem[\protect\citeauthoryear{{Ben{\'\i}tez-Llambay}}{{Ben{\'\i}tez-Llambay}}{2017}]{BenitezLlambay2017}
{Ben{\'\i}tez-Llambay} A.,  2017, {Py-SPHViewer: Cosmological simulations using
  Smoothed Particle Hydrodynamics}, Astrophysics Source Code Library, record
  ascl:1712.003 (\mn@eprint {ascl} {1712.003})

\bibitem[\protect\citeauthoryear{{Bose} et~al.,}{{Bose}
  et~al.}{2019}]{Bose2019}
{Bose} S.,  et~al., 2019, \mn@doi [\mnras] {10.1093/mnras/stz1168}, \href
  {https://ui.adsabs.harvard.edu/abs/2019MNRAS.486.4790B} {486, 4790}

\bibitem[\protect\citeauthoryear{{Brook} et~al.,}{{Brook}
  et~al.}{2011}]{Brook2011}
{Brook} C.~B.,  et~al., 2011, \mn@doi [\mnras]
  {10.1111/j.1365-2966.2011.18545.x}, \href
  {https://ui.adsabs.harvard.edu/abs/2011MNRAS.415.1051B} {415, 1051}

\bibitem[\protect\citeauthoryear{{Bullock}, {Dekel}, {Kolatt}, {Kravtsov},
  {Klypin}, {Porciani}  \& {Primack}}{{Bullock} et~al.}{2001}]{Bullock2001}
{Bullock} J.~S.,  {Dekel} A.,  {Kolatt} T.~S.,  {Kravtsov} A.~V.,  {Klypin}
  A.~A.,  {Porciani} C.,   {Primack} J.~R.,  2001, \mn@doi [\apj]
  {10.1086/321477}, \href
  {https://ui.adsabs.harvard.edu/abs/2001ApJ...555..240B} {555, 240}

\bibitem[\protect\citeauthoryear{{Celiz}, {Navarro}, {Abadi}  \&
  {Springel}}{{Celiz} et~al.}{2025}]{Celiz2025}
{Celiz} B.~M.,  {Navarro} J.~F.,  {Abadi} M.~G.,   {Springel} V.,  2025,
  \mn@doi [\aap] {10.1051/0004-6361/202554847}, \href
  {https://ui.adsabs.harvard.edu/abs/2025A&A...699A..12C} {699, A12}

\bibitem[\protect\citeauthoryear{{Cenci}, {Feldmann}, {Gensior}, {Bullock},
  {Moreno}, {Bassini}  \& {Bernardini}}{{Cenci} et~al.}{2024}]{Cenci2024}
{Cenci} E.,  {Feldmann} R.,  {Gensior} J.,  {Bullock} J.~S.,  {Moreno} J.,
  {Bassini} L.,   {Bernardini} M.,  2024, \mn@doi [\apjl]
  {10.3847/2041-8213/ad1ffb}, \href
  {https://ui.adsabs.harvard.edu/abs/2024ApJ...961L..40C} {961, L40}

\bibitem[\protect\citeauthoryear{{Chan}, {Kere{\v{s}}}, {Wetzel}, {Hopkins},
  {Faucher-Gigu{\`e}re}, {El-Badry}, {Garrison-Kimmel}  \&
  {Boylan-Kolchin}}{{Chan} et~al.}{2018}]{Chan2018}
{Chan} T.~K.,  {Kere{\v{s}}} D.,  {Wetzel} A.,  {Hopkins} P.~F.,
  {Faucher-Gigu{\`e}re} C.~A.,  {El-Badry} K.,  {Garrison-Kimmel} S.,
  {Boylan-Kolchin} M.,  2018, \mn@doi [\mnras] {10.1093/mnras/sty1153}, \href
  {https://ui.adsabs.harvard.edu/abs/2018MNRAS.478..906C} {478, 906}

\bibitem[\protect\citeauthoryear{{Chan}, {Kere{\v{s}}}, {Hopkins}, {Quataert},
  {Su}, {Hayward}  \& {Faucher-Gigu{\`e}re}}{{Chan} et~al.}{2019}]{Chan2019_CR}
{Chan} T.~K.,  {Kere{\v{s}}} D.,  {Hopkins} P.~F.,  {Quataert} E.,  {Su} K.~Y.,
   {Hayward} C.~C.,   {Faucher-Gigu{\`e}re} C.~A.,  2019, \mn@doi [\mnras]
  {10.1093/mnras/stz1895}, \href
  {https://ui.adsabs.harvard.edu/abs/2019MNRAS.488.3716C} {488, 3716}

\bibitem[\protect\citeauthoryear{{Clauwens}, {Schaye}, {Franx}  \&
  {Bower}}{{Clauwens} et~al.}{2018}]{Clauwens2018}
{Clauwens} B.,  {Schaye} J.,  {Franx} M.,   {Bower} R.~G.,  2018, \mn@doi
  [\mnras] {10.1093/mnras/sty1229}, \href
  {https://ui.adsabs.harvard.edu/abs/2018MNRAS.478.3994C} {478, 3994}

\bibitem[\protect\citeauthoryear{{Dalla Vecchia} \& {Schaye}}{{Dalla Vecchia}
  \& {Schaye}}{2008}]{DallaVecchia2008}
{Dalla Vecchia} C.,  {Schaye} J.,  2008, \mn@doi [\mnras]
  {10.1111/j.1365-2966.2008.13322.x}, \href
  {https://ui.adsabs.harvard.edu/abs/2008MNRAS.387.1431D} {387, 1431}

\bibitem[\protect\citeauthoryear{{Dekel} \& {Birnboim}}{{Dekel} \&
  {Birnboim}}{2006}]{Dekel2006}
{Dekel} A.,  {Birnboim} Y.,  2006, \mn@doi [\mnras]
  {10.1111/j.1365-2966.2006.10145.x}, \href
  {https://ui.adsabs.harvard.edu/abs/2006MNRAS.368....2D} {368, 2}

\bibitem[\protect\citeauthoryear{{El-Badry}, {Wetzel}, {Geha}, {Hopkins},
  {Kere{\v{s}}}, {Chan}  \& {Faucher-Gigu{\`e}re}}{{El-Badry}
  et~al.}{2016}]{ElBadry2016}
{El-Badry} K.,  {Wetzel} A.,  {Geha} M.,  {Hopkins} P.~F.,  {Kere{\v{s}}} D.,
  {Chan} T.~K.,   {Faucher-Gigu{\`e}re} C.-A.,  2016, \mn@doi [\apj]
  {10.3847/0004-637X/820/2/131}, \href
  {https://ui.adsabs.harvard.edu/abs/2016ApJ...820..131E} {820, 131}

\bibitem[\protect\citeauthoryear{{El-Badry} et~al.,}{{El-Badry}
  et~al.}{2018a}]{El-Badry2018}
{El-Badry} K.,  et~al., 2018a, \mn@doi [\mnras] {10.1093/mnras/stx2482}, \href
  {https://ui.adsabs.harvard.edu/abs/2018MNRAS.473.1930E} {473, 1930}

\bibitem[\protect\citeauthoryear{{El-Badry} et~al.,}{{El-Badry}
  et~al.}{2018b}]{El-Badry2018b}
{El-Badry} K.,  et~al., 2018b, \mn@doi [\mnras] {10.1093/mnras/sty730}, \href
  {https://ui.adsabs.harvard.edu/abs/2018MNRAS.477.1536E} {477, 1536}

\bibitem[\protect\citeauthoryear{{Emami}, {Siana}, {Weisz}, {Johnson}, {Ma}  \&
  {El-Badry}}{{Emami} et~al.}{2019}]{Emami2019}
{Emami} N.,  {Siana} B.,  {Weisz} D.~R.,  {Johnson} B.~D.,  {Ma} X.,
  {El-Badry} K.,  2019, \mn@doi [\apj] {10.3847/1538-4357/ab211a}, \href
  {https://ui.adsabs.harvard.edu/abs/2019ApJ...881...71E} {881, 71}

\bibitem[\protect\citeauthoryear{{Escala} et~al.,}{{Escala}
  et~al.}{2018}]{Escala2018}
{Escala} I.,  et~al., 2018, \mn@doi [\mnras] {10.1093/mnras/stx2858}, \href
  {https://ui.adsabs.harvard.edu/abs/2018MNRAS.474.2194E} {474, 2194}

\bibitem[\protect\citeauthoryear{{Faucher-Gigu{\`e}re}}{{Faucher-Gigu{\`e}re}}{2018}]{FaucherGiguere2018}
{Faucher-Gigu{\`e}re} C.-A.,  2018, \mn@doi [\mnras] {10.1093/mnras/stx2595},
  \href {https://ui.adsabs.harvard.edu/abs/2018MNRAS.473.3717F} {473, 3717}

\bibitem[\protect\citeauthoryear{{Faucher-Gigu{\`e}re}}{{Faucher-Gigu{\`e}re}}{2020}]{FaucherGiguere2020}
{Faucher-Gigu{\`e}re} C.-A.,  2020, \mn@doi [\mnras] {10.1093/mnras/staa302},
  \href {https://ui.adsabs.harvard.edu/abs/2020MNRAS.493.1614F} {493, 1614}

\bibitem[\protect\citeauthoryear{{Faucher-Gigu{\`e}re}, {Lidz}, {Zaldarriaga}
  \& {Hernquist}}{{Faucher-Gigu{\`e}re} et~al.}{2009}]{FaucherGiguere2009}
{Faucher-Gigu{\`e}re} C.-A.,  {Lidz} A.,  {Zaldarriaga} M.,   {Hernquist} L.,
  2009, \mn@doi [\apj] {10.1088/0004-637X/703/2/1416}, \href
  {https://ui.adsabs.harvard.edu/abs/2009ApJ...703.1416F} {703, 1416}

\bibitem[\protect\citeauthoryear{{Feldmann} et~al.,}{{Feldmann}
  et~al.}{2023}]{Feldmann2023}
{Feldmann} R.,  et~al., 2023, \mn@doi [\mnras] {10.1093/mnras/stad1205}, \href
  {https://ui.adsabs.harvard.edu/abs/2023MNRAS.522.3831F} {522, 3831}

\bibitem[\protect\citeauthoryear{{Flores Vel{\'a}zquez} et~al.,}{{Flores
  Vel{\'a}zquez} et~al.}{2021}]{FloresVelazquez2021}
{Flores Vel{\'a}zquez} J.~A.,  et~al., 2021, \mn@doi [\mnras]
  {10.1093/mnras/staa3893}, \href
  {https://ui.adsabs.harvard.edu/abs/2021MNRAS.501.4812F} {501, 4812}

\bibitem[\protect\citeauthoryear{{Garrison-Kimmel} et~al.,}{{Garrison-Kimmel}
  et~al.}{2018}]{GarrisonKimmel2018}
{Garrison-Kimmel} S.,  et~al., 2018, \mn@doi [\mnras] {10.1093/mnras/sty2513},
  \href {https://ui.adsabs.harvard.edu/abs/2018MNRAS.481.4133G} {481, 4133}

\bibitem[\protect\citeauthoryear{{Garrison-Kimmel} et~al.,}{{Garrison-Kimmel}
  et~al.}{2019}]{GarrisonKimmel2019}
{Garrison-Kimmel} S.,  et~al., 2019, \mn@doi [\mnras] {10.1093/mnras/stz2507},
  \href {https://ui.adsabs.harvard.edu/abs/2019MNRAS.489.4574G} {489, 4574}

\bibitem[\protect\citeauthoryear{{Guedes}, {Callegari}, {Madau}  \&
  {Mayer}}{{Guedes} et~al.}{2011}]{Guedes2011}
{Guedes} J.,  {Callegari} S.,  {Madau} P.,   {Mayer} L.,  2011, \mn@doi [\apj]
  {10.1088/0004-637X/742/2/76}, \href
  {https://ui.adsabs.harvard.edu/abs/2011ApJ...742...76G} {742, 76}

\bibitem[\protect\citeauthoryear{{Gurvich} et~al.,}{{Gurvich}
  et~al.}{2023}]{Gurvich2023}
{Gurvich} A.~B.,  et~al., 2023, \mn@doi [\mnras] {10.1093/mnras/stac3712},
  \href {https://ui.adsabs.harvard.edu/abs/2023MNRAS.519.2598G} {519, 2598}

\bibitem[\protect\citeauthoryear{{Hafen} et~al.,}{{Hafen}
  et~al.}{2022}]{Hafen2022}
{Hafen} Z.,  et~al., 2022, \mn@doi [\mnras] {10.1093/mnras/stac1603}, \href
  {https://ui.adsabs.harvard.edu/abs/2022MNRAS.514.5056H} {514, 5056}

\bibitem[\protect\citeauthoryear{{Hahn} \& {Abel}}{{Hahn} \&
  {Abel}}{2011}]{music2011}
{Hahn} O.,  {Abel} T.,  2011, \mn@doi [\mnras]
  {10.1111/j.1365-2966.2011.18820.x}, \href
  {https://ui.adsabs.harvard.edu/abs/2011MNRAS.415.2101H} {415, 2101}

\bibitem[\protect\citeauthoryear{{Helmi}, {Sales}, {Starkenburg},
  {Starkenburg}, {Vera-Ciro}, {De Lucia}  \& {Li}}{{Helmi}
  et~al.}{2012}]{Helmi2012}
{Helmi} A.,  {Sales} L.~V.,  {Starkenburg} E.,  {Starkenburg} T.~K.,
  {Vera-Ciro} C.~A.,  {De Lucia} G.,   {Li} Y.~S.,  2012, \mn@doi [\apjl]
  {10.1088/2041-8205/758/1/L5}, \href
  {https://ui.adsabs.harvard.edu/abs/2012ApJ...758L...5H} {758, L5}

\bibitem[\protect\citeauthoryear{{Hopkins}}{{Hopkins}}{2015}]{Hopkins2015}
{Hopkins} P.~F.,  2015, \mn@doi [\mnras] {10.1093/mnras/stv195}, \href
  {https://ui.adsabs.harvard.edu/abs/2015MNRAS.450...53H} {450, 53}

\bibitem[\protect\citeauthoryear{{Hopkins}}{{Hopkins}}{2017}]{Hopkins2017_MD}
{Hopkins} P.~F.,  2017, \mn@doi [\mnras] {10.1093/mnras/stw3306}, \href
  {https://ui.adsabs.harvard.edu/abs/2017MNRAS.466.3387H} {466, 3387}

\bibitem[\protect\citeauthoryear{{Hopkins}, {Kere{\v{s}}}, {O{\~n}orbe},
  {Faucher-Gigu{\`e}re}, {Quataert}, {Murray}  \& {Bullock}}{{Hopkins}
  et~al.}{2014}]{Hopkins2014}
{Hopkins} P.~F.,  {Kere{\v{s}}} D.,  {O{\~n}orbe} J.,  {Faucher-Gigu{\`e}re}
  C.-A.,  {Quataert} E.,  {Murray} N.,   {Bullock} J.~S.,  2014, \mn@doi
  [\mnras] {10.1093/mnras/stu1738}, \href
  {https://ui.adsabs.harvard.edu/abs/2014MNRAS.445..581H} {445, 581}

\bibitem[\protect\citeauthoryear{{Hopkins} et~al.,}{{Hopkins}
  et~al.}{2018a}]{Hopkins2018b}
{Hopkins} P.~F.,  et~al., 2018a, \mn@doi [\mnras] {10.1093/mnras/sty674}, \href
  {https://ui.adsabs.harvard.edu/abs/2018MNRAS.477.1578H} {477, 1578}

\bibitem[\protect\citeauthoryear{{Hopkins} et~al.,}{{Hopkins}
  et~al.}{2018b}]{Hopkins2018}
{Hopkins} P.~F.,  et~al., 2018b, \mn@doi [\mnras] {10.1093/mnras/sty1690},
  \href {https://ui.adsabs.harvard.edu/abs/2018MNRAS.480..800H} {480, 800}

\bibitem[\protect\citeauthoryear{{Hopkins}, {Grudi{\'c}}, {Wetzel},
  {Kere{\v{s}}}, {Faucher-Gigu{\`e}re}, {Ma}, {Murray}  \& {Butcher}}{{Hopkins}
  et~al.}{2020a}]{Hopkins2020a}
{Hopkins} P.~F.,  {Grudi{\'c}} M.~Y.,  {Wetzel} A.,  {Kere{\v{s}}} D.,
  {Faucher-Gigu{\`e}re} C.-A.,  {Ma} X.,  {Murray} N.,   {Butcher} N.,  2020a,
  \mn@doi [\mnras] {10.1093/mnras/stz3129}, \href
  {https://ui.adsabs.harvard.edu/abs/2020MNRAS.491.3702H} {491, 3702}

\bibitem[\protect\citeauthoryear{{Hopkins} et~al.,}{{Hopkins}
  et~al.}{2020b}]{Hopkins2020b}
{Hopkins} P.~F.,  et~al., 2020b, \mn@doi [\mnras] {10.1093/mnras/stz3321},
  \href {https://ui.adsabs.harvard.edu/abs/2020MNRAS.492.3465H} {492, 3465}

\bibitem[\protect\citeauthoryear{{Hopkins} et~al.,}{{Hopkins}
  et~al.}{2023}]{Hopkins2023}
{Hopkins} P.~F.,  et~al., 2023, \mn@doi [\mnras] {10.1093/mnras/stad1902},
  \href {https://ui.adsabs.harvard.edu/abs/2023MNRAS.525.2241H} {525, 2241}

\bibitem[\protect\citeauthoryear{{Jahn}, {Sales}, {Wetzel}, {Boylan-Kolchin},
  {Chan}, {El-Badry}, {Lazar}  \& {Bullock}}{{Jahn} et~al.}{2019}]{Jahn2019}
{Jahn} E.~D.,  {Sales} L.~V.,  {Wetzel} A.,  {Boylan-Kolchin} M.,  {Chan}
  T.~K.,  {El-Badry} K.,  {Lazar} A.,   {Bullock} J.~S.,  2019, \mn@doi
  [\mnras] {10.1093/mnras/stz2457}, \href
  {https://ui.adsabs.harvard.edu/abs/2019MNRAS.489.5348J} {489, 5348}

\bibitem[\protect\citeauthoryear{{Johnson}, {Hunter}, {Kamphuis}  \&
  {Wang}}{{Johnson} et~al.}{2017}]{Johnson2017}
{Johnson} M.~C.,  {Hunter} D.~A.,  {Kamphuis} P.,   {Wang} J.,  2017, \mn@doi
  [\mnras] {10.1093/mnrasl/slw203}, \href
  {https://ui.adsabs.harvard.edu/abs/2017MNRAS.465L..49J} {465, L49}

\bibitem[\protect\citeauthoryear{{Kaufmann}, {Wheeler}  \&
  {Bullock}}{{Kaufmann} et~al.}{2007}]{Kaufman2007}
{Kaufmann} T.,  {Wheeler} C.,   {Bullock} J.~S.,  2007, \mn@doi [\mnras]
  {10.1111/j.1365-2966.2007.12436.x}, \href
  {https://ui.adsabs.harvard.edu/abs/2007MNRAS.382.1187K} {382, 1187}

\bibitem[\protect\citeauthoryear{{Keller}, {Wadsley}, {Benincasa}  \&
  {Couchman}}{{Keller} et~al.}{2014}]{Keller2014}
{Keller} B.~W.,  {Wadsley} J.,  {Benincasa} S.~M.,   {Couchman} H.~M.~P.,
  2014, \mn@doi [\mnras] {10.1093/mnras/stu1058}, \href
  {https://ui.adsabs.harvard.edu/abs/2014MNRAS.442.3013K} {442, 3013}

\bibitem[\protect\citeauthoryear{{Kelvin} et~al.,}{{Kelvin}
  et~al.}{2012}]{Kelvin2012}
{Kelvin} L.~S.,  et~al., 2012, \mn@doi [\mnras]
  {10.1111/j.1365-2966.2012.20355.x}, \href
  {https://ui.adsabs.harvard.edu/abs/2012MNRAS.421.1007K} {421, 1007}

\bibitem[\protect\citeauthoryear{{Kim}, {Ostriker}  \& {Raileanu}}{{Kim}
  et~al.}{2017}]{Kim2017}
{Kim} C.-G.,  {Ostriker} E.~C.,   {Raileanu} R.,  2017, \mn@doi [\apj]
  {10.3847/1538-4357/834/1/25}, \href
  {https://ui.adsabs.harvard.edu/abs/2017ApJ...834...25K} {834, 25}

\bibitem[\protect\citeauthoryear{{Klein}, {Bullock}, {Moreno}, {Mercado},
  {Hopkins}, {Cochrane}  \& {Benavides}}{{Klein} et~al.}{2024}]{Klein2024}
{Klein} C.,  {Bullock} J.~S.,  {Moreno} J.,  {Mercado} F.~J.,  {Hopkins} P.~F.,
   {Cochrane} R.~K.,   {Benavides} J.~A.,  2024, \mn@doi [\mnras]
  {10.1093/mnras/stae1505}, \href
  {https://ui.adsabs.harvard.edu/abs/2024MNRAS.532..538K} {532, 538}

\bibitem[\protect\citeauthoryear{{Klein} et~al.,}{{Klein}
  et~al.}{2025}]{Klein2025}
{Klein} C.,  et~al., 2025, \mn@doi [arXiv e-prints]
  {10.48550/arXiv.2503.05612}, \href
  {https://ui.adsabs.harvard.edu/abs/2025arXiv250305612K} {p. arXiv:2503.05612}

\bibitem[\protect\citeauthoryear{{Knollmann} \& {Knebe}}{{Knollmann} \&
  {Knebe}}{2009}]{AHF}
{Knollmann} S.~R.,  {Knebe} A.,  2009, \mn@doi [\apjs]
  {10.1088/0067-0049/182/2/608}, \href
  {https://ui.adsabs.harvard.edu/abs/2009ApJS..182..608K} {182, 608}

\bibitem[\protect\citeauthoryear{{Lange} et~al.,}{{Lange}
  et~al.}{2016}]{Lange2016}
{Lange} R.,  et~al., 2016, \mn@doi [\mnras] {10.1093/mnras/stw1495}, \href
  {https://ui.adsabs.harvard.edu/abs/2016MNRAS.462.1470L} {462, 1470}

\bibitem[\protect\citeauthoryear{{Li}, {Li}, {Cui}, {Marinacci}, {Sales},
  {Vogelsberger}  \& {Torrey}}{{Li} et~al.}{2024}]{Chengzhe2024}
{Li} C.,  {Li} H.,  {Cui} W.,  {Marinacci} F.,  {Sales} L.~V.,  {Vogelsberger}
  M.,   {Torrey} P.,  2024, \mn@doi [\mnras] {10.1093/mnras/stae797}, \href
  {https://ui.adsabs.harvard.edu/abs/2024MNRAS.529.4073L} {529, 4073}

\bibitem[\protect\citeauthoryear{{Marinacci}, {Pakmor}  \&
  {Springel}}{{Marinacci} et~al.}{2014}]{Marinacci2014}
{Marinacci} F.,  {Pakmor} R.,   {Springel} V.,  2014, \mn@doi [\mnras]
  {10.1093/mnras/stt2003}, \href
  {https://ui.adsabs.harvard.edu/abs/2014MNRAS.437.1750M} {437, 1750}

\bibitem[\protect\citeauthoryear{{Martizzi}}{{Martizzi}}{2020}]{Martizzi2020}
{Martizzi} D.,  2020, \mn@doi [\mnras] {10.1093/mnras/stz3419}, \href
  {https://ui.adsabs.harvard.edu/abs/2020MNRAS.492...79M} {492, 79}

\bibitem[\protect\citeauthoryear{{McCluskey}, {Wetzel}, {Loebman}, {Moreno},
  {Faucher-Gigu{\`e}re}  \& {Hopkins}}{{McCluskey}
  et~al.}{2024}]{McCluskey2024}
{McCluskey} F.,  {Wetzel} A.,  {Loebman} S.~R.,  {Moreno} J.,
  {Faucher-Gigu{\`e}re} C.-A.,   {Hopkins} P.~F.,  2024, \mn@doi [\mnras]
  {10.1093/mnras/stad3547}, \href
  {https://ui.adsabs.harvard.edu/abs/2024MNRAS.527.6926M} {527, 6926}

\bibitem[\protect\citeauthoryear{{McCluskey}, {Wetzel}, {Loebman}  \&
  {Moreno}}{{McCluskey} et~al.}{2025}]{McCluskey2025}
{McCluskey} F.,  {Wetzel} A.,  {Loebman} S.,   {Moreno} J.,  2025, \mn@doi
  [arXiv e-prints] {10.48550/arXiv.2506.11840}, \href
  {https://ui.adsabs.harvard.edu/abs/2025arXiv250611840M} {p. arXiv:2506.11840}

\bibitem[\protect\citeauthoryear{{Mercado} et~al.,}{{Mercado}
  et~al.}{2025}]{Mercado2025}
{Mercado} F.~J.,  et~al., 2025, \mn@doi [\apj] {10.3847/1538-4357/adbf07},
  \href {https://ui.adsabs.harvard.edu/abs/2025ApJ...983...93M} {983, 93}

\bibitem[\protect\citeauthoryear{{Moreno} et~al.,}{{Moreno}
  et~al.}{2022}]{Moreno2022}
{Moreno} J.,  et~al., 2022, \mn@doi [Nature Astronomy]
  {10.1038/s41550-021-01598-4}, \href
  {https://ui.adsabs.harvard.edu/abs/2022NatAs...6..496M} {6, 496}

\bibitem[\protect\citeauthoryear{{Munshi}, {Brooks}, {Applebaum},
  {Christensen}, {Quinn}  \& {Sligh}}{{Munshi} et~al.}{2021}]{Munshi2021}
{Munshi} F.,  {Brooks} A.~M.,  {Applebaum} E.,  {Christensen} C.~R.,  {Quinn}
  T.,   {Sligh} S.,  2021, \mn@doi [\apj] {10.3847/1538-4357/ac0db6}, \href
  {https://ui.adsabs.harvard.edu/abs/2021ApJ...923...35M} {923, 35}

\bibitem[\protect\citeauthoryear{{Oh} et~al.,}{{Oh} et~al.}{2015}]{Oh2015}
{Oh} S.-H.,  et~al., 2015, \mn@doi [\aj] {10.1088/0004-6256/149/6/180}, \href
  {https://ui.adsabs.harvard.edu/abs/2015AJ....149..180O} {149, 180}

\bibitem[\protect\citeauthoryear{{Orr}, {Fielding}, {Hayward}  \&
  {Burkhart}}{{Orr} et~al.}{2022}]{Orr2022}
{Orr} M.~E.,  {Fielding} D.~B.,  {Hayward} C.~C.,   {Burkhart} B.,  2022,
  \mn@doi [\apj] {10.3847/1538-4357/ac6c26}, \href
  {https://ui.adsabs.harvard.edu/abs/2022ApJ...932...88O} {932, 88}

\bibitem[\protect\citeauthoryear{{Planck Collaboration} et~al.,}{{Planck
  Collaboration} et~al.}{2016}]{Planck2016}
{Planck Collaboration} et~al., 2016, \mn@doi [\aap]
  {10.1051/0004-6361/201525830}, \href
  {https://ui.adsabs.harvard.edu/abs/2016A&A...594A..13P} {594, A13}

\bibitem[\protect\citeauthoryear{{Rodriguez-Gomez} et~al.,}{{Rodriguez-Gomez}
  et~al.}{2017}]{RodriguezGomez2017}
{Rodriguez-Gomez} V.,  et~al., 2017, \mn@doi [\mnras] {10.1093/mnras/stx305},
  \href {https://ui.adsabs.harvard.edu/abs/2017MNRAS.467.3083R} {467, 3083}

\bibitem[\protect\citeauthoryear{{Roychowdhury}, {Chengalur}, {Karachentsev}
  \& {Kaisina}}{{Roychowdhury} et~al.}{2013}]{Roychowdhury2013}
{Roychowdhury} S.,  {Chengalur} J.~N.,  {Karachentsev} I.~D.,   {Kaisina}
  E.~I.,  2013, \mn@doi [\mnras] {10.1093/mnrasl/slt123}, \href
  {https://ui.adsabs.harvard.edu/abs/2013MNRAS.436L.104R} {436, L104}

\bibitem[\protect\citeauthoryear{{Sales}, {Navarro}, {Theuns}, {Schaye},
  {White}, {Frenk}, {Crain}  \& {Dalla Vecchia}}{{Sales}
  et~al.}{2012}]{Sales2012}
{Sales} L.~V.,  {Navarro} J.~F.,  {Theuns} T.,  {Schaye} J.,  {White} S. D.~M.,
   {Frenk} C.~S.,  {Crain} R.~A.,   {Dalla Vecchia} C.,  2012, \mn@doi [\mnras]
  {10.1111/j.1365-2966.2012.20975.x}, \href
  {https://ui.adsabs.harvard.edu/abs/2012MNRAS.423.1544S} {423, 1544}

\bibitem[\protect\citeauthoryear{{Santos-Santos} et~al.,}{{Santos-Santos}
  et~al.}{2020}]{SantosSantos2020}
{Santos-Santos} I. M.~E.,  et~al., 2020, \mn@doi [\mnras]
  {10.1093/mnras/staa1072}, \href
  {https://ui.adsabs.harvard.edu/abs/2020MNRAS.495...58S} {495, 58}

\bibitem[\protect\citeauthoryear{{Scannapieco}, {White}, {Springel}  \&
  {Tissera}}{{Scannapieco} et~al.}{2009}]{Scannapieco2009}
{Scannapieco} C.,  {White} S. D.~M.,  {Springel} V.,   {Tissera} P.~B.,  2009,
  \mn@doi [\mnras] {10.1111/j.1365-2966.2009.14764.x}, \href
  {https://ui.adsabs.harvard.edu/abs/2009MNRAS.396..696S} {396, 696}

\bibitem[\protect\citeauthoryear{{Simons}, {Kassin}, {Weiner}, {Heckman},
  {Lee}, {Lotz}, {Peth}  \& {Tchernyshyov}}{{Simons} et~al.}{2015}]{Simons2015}
{Simons} R.~C.,  {Kassin} S.~A.,  {Weiner} B.~J.,  {Heckman} T.~M.,  {Lee}
  J.~C.,  {Lotz} J.~M.,  {Peth} M.,   {Tchernyshyov} K.,  2015, \mn@doi
  [\mnras] {10.1093/mnras/stv1298}, \href
  {https://ui.adsabs.harvard.edu/abs/2015MNRAS.452..986S} {452, 986}

\bibitem[\protect\citeauthoryear{{Springel}}{{Springel}}{2005}]{Springel2005}
{Springel} V.,  2005, \mn@doi [\mnras] {10.1111/j.1365-2966.2005.09655.x},
  \href {https://ui.adsabs.harvard.edu/abs/2005MNRAS.364.1105S} {364, 1105}

\bibitem[\protect\citeauthoryear{{Springel} \& {Hernquist}}{{Springel} \&
  {Hernquist}}{2003}]{Springel2003}
{Springel} V.,  {Hernquist} L.,  2003, \mn@doi [\mnras]
  {10.1046/j.1365-8711.2003.06206.x}, \href
  {https://ui.adsabs.harvard.edu/abs/2003MNRAS.339..289S} {339, 289}

\bibitem[\protect\citeauthoryear{{Stern} et~al.,}{{Stern}
  et~al.}{2021}]{Stern2021}
{Stern} J.,  et~al., 2021, \mn@doi [\apj] {10.3847/1538-4357/abd776}, \href
  {https://ui.adsabs.harvard.edu/abs/2021ApJ...911...88S} {911, 88}

\bibitem[\protect\citeauthoryear{{Stern}, {Fielding}, {Hafen}, {Su}, {Naor},
  {Faucher-Gigu{\`e}re}, {Quataert}  \& {Bullock}}{{Stern}
  et~al.}{2024}]{Stern2024}
{Stern} J.,  {Fielding} D.,  {Hafen} Z.,  {Su} K.-Y.,  {Naor} N.,
  {Faucher-Gigu{\`e}re} C.-A.,  {Quataert} E.,   {Bullock} J.,  2024, \mn@doi
  [\mnras] {10.1093/mnras/stae824}, \href
  {https://ui.adsabs.harvard.edu/abs/2024MNRAS.530.1711S} {530, 1711}

\bibitem[\protect\citeauthoryear{{Suarez}, {Pontzen}, {Peiris}, {Slyz}  \&
  {Devriendt}}{{Suarez} et~al.}{2016}]{Suarez2016}
{Suarez} T.,  {Pontzen} A.,  {Peiris} H.~V.,  {Slyz} A.,   {Devriendt} J.,
  2016, \mn@doi [\mnras] {10.1093/mnras/stw1670}, \href
  {https://ui.adsabs.harvard.edu/abs/2016MNRAS.462..994S} {462, 994}

\bibitem[\protect\citeauthoryear{{Tacchella} et~al.,}{{Tacchella}
  et~al.}{2019}]{Tacchella2019}
{Tacchella} S.,  et~al., 2019, \mn@doi [\mnras] {10.1093/mnras/stz1657}, \href
  {https://ui.adsabs.harvard.edu/abs/2019MNRAS.487.5416T} {487, 5416}

\bibitem[\protect\citeauthoryear{{{\"U}bler}, {Naab}, {Oser}, {Aumer}, {Sales}
  \& {White}}{{{\"U}bler} et~al.}{2014}]{Ubler2014}
{{\"U}bler} H.,  {Naab} T.,  {Oser} L.,  {Aumer} M.,  {Sales} L.~V.,   {White}
  S. D.~M.,  2014, \mn@doi [\mnras] {10.1093/mnras/stu1275}, \href
  {https://ui.adsabs.harvard.edu/abs/2014MNRAS.443.2092U} {443, 2092}

\bibitem[\protect\citeauthoryear{{Wetzel}, {Hopkins}, {Kim},
  {Faucher-Gigu{\`e}re}, {Kere{\v{s}}}  \& {Quataert}}{{Wetzel}
  et~al.}{2016}]{Wetzel2016}
{Wetzel} A.~R.,  {Hopkins} P.~F.,  {Kim} J.-h.,  {Faucher-Gigu{\`e}re} C.-A.,
  {Kere{\v{s}}} D.,   {Quataert} E.,  2016, \mn@doi [\apjl]
  {10.3847/2041-8205/827/2/L23}, \href
  {https://ui.adsabs.harvard.edu/abs/2016ApJ...827L..23W} {827, L23}

\bibitem[\protect\citeauthoryear{{Wetzel} et~al.,}{{Wetzel}
  et~al.}{2023}]{Wetzel2023}
{Wetzel} A.,  et~al., 2023, \mn@doi [\apjs] {10.3847/1538-4365/acb99a}, \href
  {https://ui.adsabs.harvard.edu/abs/2023ApJS..265...44W} {265, 44}

\bibitem[\protect\citeauthoryear{{Yu} et~al.,}{{Yu} et~al.}{2021}]{Yu2021}
{Yu} S.,  et~al., 2021, \mn@doi [\mnras] {10.1093/mnras/stab1339}, \href
  {https://ui.adsabs.harvard.edu/abs/2021MNRAS.505..889Y} {505, 889}

\bibitem[\protect\citeauthoryear{{Yu} et~al.,}{{Yu} et~al.}{2023}]{Yu2023}
{Yu} S.,  et~al., 2023, \mn@doi [\mnras] {10.1093/mnras/stad1806}, \href
  {https://ui.adsabs.harvard.edu/abs/2023MNRAS.523.6220Y} {523, 6220}

\bibitem[\protect\citeauthoryear{{Zeng}, {Wang}, {Gao}  \& {Yang}}{{Zeng}
  et~al.}{2024}]{Zeng2024}
{Zeng} G.,  {Wang} L.,  {Gao} L.,   {Yang} H.,  2024, \mn@doi [\mnras]
  {10.1093/mnras/stae1651}, \href
  {https://ui.adsabs.harvard.edu/abs/2024MNRAS.532.2558Z} {532, 2558}

\bibitem[\protect\citeauthoryear{{Zhang} et~al.,}{{Zhang}
  et~al.}{2024}]{Zhang2024}
{Zhang} E.,  et~al., 2024, \mn@doi [\apj] {10.3847/1538-4357/ad7f57}, \href
  {https://ui.adsabs.harvard.edu/abs/2024ApJ...975..229Z} {975, 229}

\makeatother
\end{thebibliography}







\bsp	
\label{lastpage}
\end{document}